\documentclass[submission,Phys]{SciPost}

\usepackage[utf8]{inputenc} 
\usepackage[T1]{fontenc} 	
\usepackage[english]{babel} 

\usepackage[bitstream-charter]{mathdesign}
\usepackage{geometry} 		
\usepackage{amsmath} 		
\usepackage{amsthm} 		
\usepackage{mathtools} 		
\usepackage{float} 			
\usepackage{graphicx} 		
\usepackage{tabularx} 		
\usepackage{booktabs} 		
\usepackage{color, xcolor} 	
\usepackage{pdfpages} 		
\usepackage{extarrows} 		
\usepackage{multirow} 		
\usepackage{multicol} 		
\usepackage{caption} 		
\usepackage{subcaption} 	
\usepackage{enumitem} 		
\usepackage{setspace} 		
\usepackage{xspace} 		
\usepackage{ragged2e} 		
\usepackage{stackrel} 		
\usepackage{tikz} 			
\usepackage{braket} 		
\usepackage{bm} 			
\usepackage{tensor} 		
\usepackage{slashed} 		
\usepackage{siunitx} 		
\usepackage{lastpage} 		
\usepackage{cite} 			
\usepackage[normalem]{ulem} 
\usepackage{fontawesome} 	
\usepackage{tocloft} 		
\usepackage{titlesec} 		
\usepackage{doi} 			
\usepackage{hyperref} 		
\usepackage[most]{tcolorbox} 					
\usepackage[nameinlink, capitalize]{cleveref} 	
\usepackage[nottoc, notlot, notlof]{tocbibind} 	
\usepackage[ruled, vlined]{algorithm2e} 		

\hypersetup{
	pdftitle={Loop Amplitudes from Precision Networks},
	pdfauthor={Simon Badger, Anja Butter, Michel Luchmann, Sebastian Pitz and Tilman Plehn},
	colorlinks=true, 			
	linkcolor={red!50!black}, 	
	citecolor={blue!50!black}, 	
	urlcolor={blue!80!black} 	
} 

\DeclareSymbolFont{usualmathcal}{OMS}{cmsy}{m}{n}
\DeclareSymbolFontAlphabet{\mathcal}{usualmathcal}

\SetArgSty{textnormal}
\SetKwComment{Comment}{{\small\#}~}{}
\SetCommentSty{mycommfont}

\setitemize{itemsep=0pt, parsep=0pt} 				
\setenumerate{itemsep=0pt, parsep=0pt} 				
\setitemize{itemsep=2pt,topsep=2pt,parsep=0pt,partopsep=0pt,leftmargin=*}
\setenumerate{itemsep=0pt,topsep=2pt,parsep=0pt,partopsep=0pt,labelindent=3pt,leftmargin=*}
\newcommand{\kl}{\text{KL}}

\newcommand{\XXLangle}{\biggl\langle}
\newcommand{\XXRangle}{\biggr\rangle}

\newcommand{\qqquad}{\qquad\quad}
\newcommand{\qqqquad}{\qquad\qquad}

\newcommand\one{\leavevmode\hbox{\small1\normalsize\kern-.33em1}}

\newcommand{\loss}{\mathcal{L}} 	

\newcommand{\arXiv}[2][]{%
	\ifthenelse{\equal{#1}{}}%
	{\href{http://arxiv.org/abs/#2}{arXiv:#2}}%
	{\href{http://arxiv.org/abs/#2}{arXiv:#2~[#1]}}}

\newcommand{\gev}{\text{GeV}}

\def\slashchar#1{\setbox0=\hbox{$#1$}           
   \dimen0=\wd0                                 
   \setbox1=\hbox{/} \dimen1=\wd1               
   \ifdim\dimen0>\dimen1                        
      \rlap{\hbox to \dimen0{\hfil/\hfil}}      
      #1                                        
   \else                                        
      \rlap{\hbox to \dimen1{\hfil$#1$\hfil}}   
      /                                         
   \fi}

\graphicspath{{./figs/}}                

\begin{document}

\begin{center}{\Large \textbf{
Loop Amplitudes from Precision Networks
}}\end{center}

\begin{center}
  Simon Badger\textsuperscript{1},
  Anja Butter\textsuperscript{2,3},
  Michel Luchmann\textsuperscript{2},
  Sebastian Pitz\textsuperscript{2} and 
  Tilman Plehn\textsuperscript{2}
\end{center}

\begin{center}
  {\bf 1} Physics Department, Torino University and INFN Torino, Torino, Italy \\
  {\bf 2} Institut f\"ur Theoretische Physik, Universit\"at Heidelberg, Germany\\
  {\bf 3} LPNHE, Sorbonne Universit\'e, Universit\'e Paris Cit\'e, CNRS/IN2P3, Paris, France
\end{center}

\begin{center}
\today
\end{center}

\section*{Abstract}
         {\bf Evaluating loop amplitudes is a time-consuming part of
           LHC event generation. For di-photon production with jets we
           show that simple, Bayesian networks can learn such
           amplitudes and model their uncertainties reliably. A
           boosted training of the Bayesian network further improves
           the uncertainty estimate and the network precision in
           critical phase space regions. In general, boosted network
           training of Bayesian networks allows us to move between
           fit-like and interpolation-like regimes of network
           training.}

\vspace{10pt}
\noindent\rule{\textwidth}{1pt}
\tableofcontents\thispagestyle{fancy}
\noindent\rule{\textwidth}{1pt}
\vspace{10pt}

\clearpage
\section{Introduction}
\label{sec:intro}

Combining our expectations of a vastly increased dataset from the
upcoming LHC runs with novel analysis methods and ever-improving
theory predictions, we are looking at exciting times for particle
physics. One of the keys to make optimal use of the LHC data is to
consistently employ modern techniques, inspired by data science and
further developed for particle physics application. Inference based on
precision predictions from first principles critically rests on the
assumption that we can provide theory predictions over the full phase space
fast, precisely, and with flexible model assumptions. To meet the
speed and precision expectations from HL-LHC we can use modern machine
learning (ML) throughout the event generation and simulation
chain~\cite{Butter:2022rso}.

A straightforward ML-task is regression of loop amplitudes,
represented as a smooth scalar function over a relatively simple phase
space. For simple ($2 \to 2$)-processes learning a non-divergent loop
amplitude does not even require deep
networks~\cite{Bishara:2019iwh} and has been achieved with conventional interpolation methods~\cite{Czakon:2008zk, Borowka:2016ehy} as well. For higher final-state
multiplicities~\cite{Danziger:2021eeg} precision turns into a
challenge, which we can try to meet by separating phase space into
finite and divergent regions~\cite{Badger:2020uow} or physics-inspired
channels combined with very large training
samples~\cite{Maitre:2021uaa}. For our di-photon benchmark process at
one-loop order current methods have shown to work well, but with
limited precision especially in challenging regions of phase
space~\cite{Aylett-Bullock:2021hmo}.

Like in many physics applications, we would like to complement
precision predictions of amplitudes with a reliable uncertainty
estimate. Amplitudes are a simple problem because the training data
consists of arbitrarily precise numerical values for well-defined
phase space points. Once the network has learned all relevant
features, we expect the leading uncertainties to reflect a possible
local sparsity of the training data.  Bayesian networks are perfectly
suited to track training-related uncertainties~\cite{bnn_early3}.  In
LHC physics they have been applied to
regression~\cite{Kasieczka:2020vlh},
classification~\cite{Bollweg:2019skg}, ensembling~\cite{Araz:2021wqm},
and generation~\cite{Bellagente:2021yyh,Butter:2021csz}. We will use
Bayesian networks as a surrogate for ML-amplitudes because they learn
amplitude values together with an uncertainty, and because we can use
their likelihood loss to improve the network training.

When we want to train a network on amplitudes over phase space, with
the additional condition that large amplitude values should be
reproduced well, we need to re-think our training strategy. While
usual NN-applications can be viewed as a non-parametric fit, we want
to precisely reproduce individual amplitudes in the spirit of an
interpolation~\cite{Chahrour:2021eiv}.  We can force the network to
reproduce certain amplitudes by boosting these amplitudes in the
training.  To decide which amplitudes need boosting, we use a Bayesian
network with its point-wise uncertainty estimate. We find that moving
freely between fit-like and the interpolation-like tasks allows us to
improve the uncertainty estimate through a loss-based boosting and the
precision though a process-specific performance boosting.

In Sec.~\ref{sec:data} we introduce our dataset and the benchmark
results, before introducing the Bayesian network in
Sec.~\ref{sec:bnn}. The improved training through the two boosting
strategies is illustrated for the $\gamma \gamma g$ channel in
Sec.~\ref{sec:boosting}. Finally, we compare a set of 1-dimension
kinematic distributions for the training data and the NN-amplitudes
including uncertainties and with the different boosting strategies,
before we provide an Outlook. In the Appendix we show the
corresponding results for the $\gamma \gamma gg$ final state, based on
the same concepts and architectures, but using a larger network and
more training data.


\section{Dataset and benchmark results}
\label{sec:data}

As an example process for our surrogate NN-amplitudes we use the
partonic one-loop process~\cite{Aylett-Bullock:2021hmo}
\begin{align}
gg \to \gamma \gamma g (g) \; ,
\label{eq:proc}
\end{align}
generated with \textsc{Sherpa}~\cite{Sherpa:2019gpd} and the \textsc{NJet} amplitude library~\cite{Badger:2012pg}\footnote{In the main body
of the paper we work with the ($2 \to 3$)-process, the corresponding results for the 4-body final state are presented in the Appendix. The interface between \textsc{NJet} and \textsc{Sherpa} is provided with Ref.~\cite{Aylett-Bullock:2021hmo} and available at \url{https://github.com/JosephPB/n3jet}.}.  We
apply a basic set of detector-inspired cuts on the partons in the
final state,
\begin{alignat}{9}
p_{T,j} &> 20~\gev
&\qqqquad
| \eta_j | &< 5
&\qqqquad 
R_{jj, j \gamma, \gamma \gamma} > 0.4 \notag \\
p_{T,\gamma} &> 40, 30~\gev
&\qqqquad
| \eta_\gamma | &< 2.37 \; .
\label{eq:cuts}
\end{alignat}
These cuts reduce the originally produced dataset from 1.1M ($2 \to
3$)-amplitudes of Ref.~\cite{Aylett-Bullock:2021hmo} to roughly 960k
amplitudes.
Each data point consists of real amplitude values as a function of the
external 4-momenta, defining a 20-dimensional phase space.  We divide
our dataset into 90k training amplitudes and 870k independent test
amplitudes acting as a high-statistics truth.\medskip

Transition amplitudes play a special role in the applications of
neural networks to LHC physics, because they can be computed as
functions of phase space with essentially arbitrary precision. The
combination of high precision with limited training data is the
challenge for the corresponding regression networks. Implicitly, it is
assumed that the NN-amplitude networks will be faster than even the
evaluation of leading-order amplitudes with state-of-the-art methods.
The main figure of merit compares the true and the NN-amplitudes for a
set of training or test data points,
\begin{align}
\Delta_j^\text{(train)} = \frac{A_{j,\text{NN}}}{A_{j,\text{train}}} - 1
\qqquad \text{or} \qqquad
\Delta_j^\text{(test)} = \frac{A_{j,\text{NN}}}{A_{j,\text{test}}} - 1 \; ,
\label{eq:def_deltas}
\end{align}
where $j$ runs over amplitude data points and we subtract $1$ compared
with the original paper~\cite{Aylett-Bullock:2021hmo}. Typical
distributions of these $\Delta$ for existing calculations come with a
width of 10\% or more for the one-jet process of
Eq.\eqref{eq:proc}~\cite{Aylett-Bullock:2021hmo}. For the tree-level process $e^+
e^- \to q\bar{q}g$ the width of the $\Delta$-distribution can be
reduced to the per-mille level, using a training dataset of 60M
amplitudes and a rather complex, physics-inspired architecture of
networks~\cite{Maitre:2021uaa}.

Our approach follows a different strategy from the physics-inspired
architectures mentioned above. We will use a relatively simple and
small network, enhanced by a Bayesian network structure, and target
the precision requirements with a new training strategy. The goal is
to show that small and hence fast networks are expressive enough to
describe a scattering amplitude over phase space. Enforcing and
controlling the required precision leads us to, essentially, an
appropriate loss function and the corresponding network training
strategy.

\section{Bayesian network}
\label{sec:bnn}

Deterministic networks, usually trained by minimizing an MSE loss
function, exhibit several weaknesses when it comes to LHC applications
and controlled precision predictions. First, they only learn the
amplitude value over phase space, without any information on if they
have learned all features and how precise their estimate is. Second,
their conceptually weak MSE loss function limits their performance.
We will show how a Bayesian network with a likelihood comes with a
whole range of conceptual and practical benefits.

\subsubsection*{Bayesian networks and uncertainties}

In contrast to standard, deterministic networks, Bayesian neural
networks (BNNs) learn distributions of network parameters or weights
$\omega$~\cite{bnn_early,bnn_early3}.  Sampling over the
weight distributions gives us an uncertainty in the network
output. At the end of this introduction we will approximate each
weight distribution by a Gaussian, which does not limit the
expressivity of a deep Bayesian network, but means that the
Bayesian network requires only twice as many parameters as its
deterministic counterpart~\cite{bnn_early3}. By definition, the
Bayesian network includes a generalized dropout and an explicit
regularization term in the loss, which stabilize the training.

\begin{figure}[b!]
  \centering
  \includegraphics[width=0.8\textwidth, trim=0cm 4.5cm 0cm 0cm]{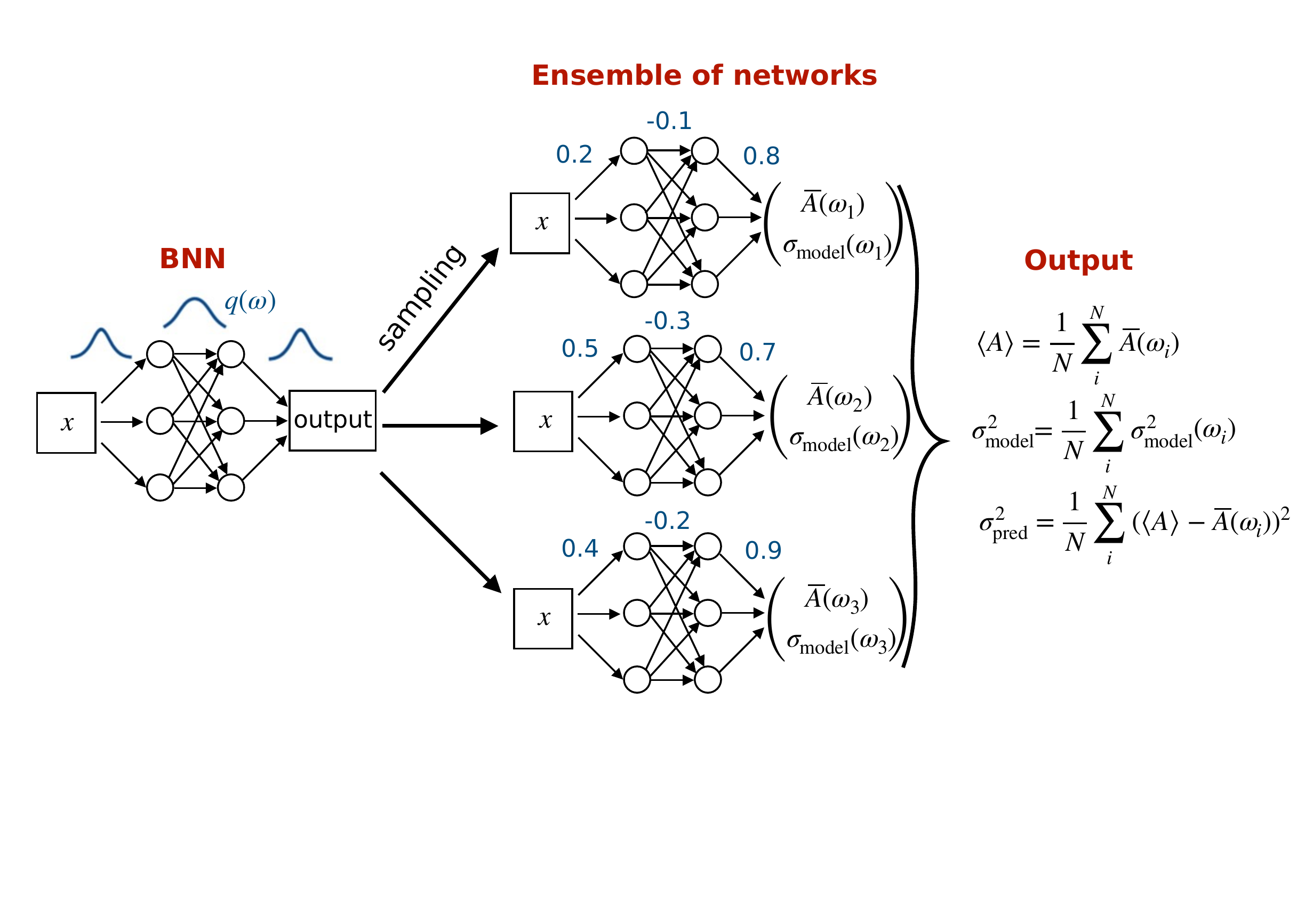}
  \caption{Illustration of the Bayesian network.}
  \label{fig:bnn}
\end{figure}

With our amplitude network we want to predict the transition amplitude
$A$ for a phase space point $x$.  If we define $p(A|x) \equiv p(A)$ as
the probability distribution for possible amplitudes at a given phase
space point $x$, and omitting the argument $x$ from now on, its mean
value is
\begin{align}
\langle \, A \, \rangle 
= \int dA \; A \; p(A)
\qquad \text{with} \qquad 
  p(A) 
  = \int d \omega \; p(A | \omega) \; p(\omega |T) \; ,
\label{eq:folding}
\end{align}
where $p(\omega|T)$ are the network weight distribution and $T$ is the
training data.  We do not know the closed form of $p(\omega | T)$, but
we can approximate it with a simpler tractable distribution $q(\omega)$:
\begin{align}
p(A)
= \int d \omega \; p(A | \omega) \; p(\omega | T)
\approx \int d \omega \; p(A | \omega) \; q(\omega) \; .
\label{eq:first_bayes}
\end{align}
This approximation leads us directly to the BNN loss function. We
implement the variational approximation as a Kullback-Leibler
divergence,
\begin{align}
\kl [q(\omega),p(\omega|T)] 
&= \int d\omega \; q(\omega) \; \log \frac{q(\omega)}{p(\omega|T)} \notag \\
&= \int d\omega \; q(\omega) \; \log \frac{q(\omega) p(T)}{p(\omega) p( T|\omega)} \notag \\
&= \kl[q(\omega),p(\omega)] 
 - \int d\omega \; q(\omega) \; \log p(T|\omega)
 + \log p(T) \int d\omega \; q(\omega) \; .
\label{eq:kl_bayes}
\end{align}
Bayes' theorem gives the corresponding networks their name. The prior
$p(\omega)$ describes the model parameters before training. The model
evidence $p(T)$ guarantees the correct normalization of $p(\omega | T
)$.  Turning Eq.\eqref{eq:kl_bayes} into a loss function we can omit
the evidence, if we enforce the normalization condition another way,
\begin{align}
\loss_\text{BNN} 
= - \int d\omega \; q(\omega) \; \log p(T|\omega) 
+ \kl[q(\omega),p(\omega)] \; .
\label{eq:loss_bayes1}
\end{align}
The combined log-likelihood $\log p(T|\omega)$ implicitly includes the
sum over all training points.\medskip

To get access to the mean and the uncertainty of the network
prediction for $A$, we exchange the two integrals in
Eq.\eqref{eq:folding} and find 
\begin{align}
\langle A \rangle 
&= \int dA  \;  d\omega \; A \; p(A | \omega, T) \; q(\omega) \notag \\
&\equiv \int d\omega \; q(\omega) \; \overline{A}(\omega)  
\qquad \text{with} \qquad
\overline{A}(\omega) 
= \int dA \; A \; p(A | \omega) \; .
\label{eq:expectations}
\end{align}
A network with perfect $x$-resolution and perfect interpolation
properties would be described by $q(\omega) =
\delta(\omega-\omega_0)$, and $p(A|\omega)$ would simply return the
one correct value for the amplitude. For noisy training data
$p(A|\omega)$ actually describes a spectrum reflecting the noisy
labels~\cite{Kasieczka:2020vlh}. In our case, the amplitudes are
exact, but the network will still not interpolate perfectly
between the sparse training data points.  Corresponding to
Eq.\eqref{eq:expectations} the variance of $A$ is
\begin{align}
\sigma_\text{tot}^2
&= \int dA \;  d \omega \; \left( A - \langle A \rangle \right)^2 \; p(A | \omega) \; q(\omega) \notag \\
&= \int d\omega \; q(\omega) \left[ 
   \int dA \; A^2 \; p(A | \omega) 
 - 2 \langle A \rangle \int dA \; A  \; p(A | \omega) 
 + \langle A \rangle^2 \int dA  \; p(A | \omega) \right] \notag \\
&= \int d\omega \; q(\omega) \left[ 
   \overline{A^2}(\omega) 
 - 2 \langle A \rangle \overline{A}(\omega) 
 + \langle A \rangle^2 \right] \notag \\
&= \int d\omega \; q(\omega) \left[ 
   \overline{A^2}(\omega) - \overline{A}(\omega)^2
 + \left( \overline{A}(\omega) - \langle A \rangle \right)^2 \right] 
\equiv \sigma_\text{model}^2 + \sigma_\text{pred}^2 \; ,
\label{eq:def_sigmas}
\end{align}
where $\overline{A^2}(\omega)$ is defined in analogy to
$\overline{A}(\omega)$ in Eq.\eqref{eq:expectations}.  This defines
two contributions to the uncertainty.  First, $\sigma_\text{pred}$ is
defined in terms of the $\omega$-integrated expectation value $\langle
A \rangle$
\begin{align}
\sigma_\text{pred}^2
&= \int d\omega \; q(\omega) \;
   \Big[ \overline{A}(\omega) - \langle A \rangle \Big]^2 \; .
\label{eq:sig_pred}
\end{align}
It vanishes in the limit of perfect training (with infinite training data), 
$q(\omega)\to \delta(\omega -\omega_0)$. In that sense it represents 
a statistical uncertainty, and
for a well-trained network we expect it to become small. In contrast,
$\sigma_\text{model}$ already occurs without sampling the network
parameters,
\begin{align}
\sigma_\text{model}^2 
\equiv \langle \sigma_\text{model}(\omega)^2 \rangle 
=& \int d\omega \; q(\omega) \; \sigma_\text{model}(\omega)^2 \notag \\
=& \int d\omega \; q(\omega) \; \Big[
   \overline{A^2}(\omega) - \overline{A}(\omega)^2 \Big] \; .
\label{eq:sig_model}
\end{align}
It will be induced by limited training data, but in the limit of
perfect training it approaches a plateau, accounting for a
non-deterministic or stochastic label, limited expressivity of the
network, not-so-smart choices of hyperparameters etc, in the sense of
a systematic uncertainty. To avoid mis-understanding we refer to it as
a model-related uncertainty rather than the usual
$\sigma_\text{stoch}$ in case our data is non-stochastic, like the
amplitudes in this application.

To understand these two uncertainty measures better, we can read
Eqs.\eqref{eq:expectations} and~\eqref{eq:sig_model} as a sampling of
$\overline{A}(\omega)$ and $\sigma_\text{model}(\omega)^2$ over a
Gaussian network parameter distribution $q(\omega)$. This sampling
uses the network-encoded amplitude and uncertainty values over phase
space $x$ and network parameter space $\omega$,
\begin{align}
\text{BNN}: x, \omega \; \to \; 
\begin{pmatrix}
\overline{A}(\omega)\\
\sigma_\text{model}(\omega)
\end{pmatrix}  \; .
\label{eq:bnn_output}
\end{align}
%
\medskip

Until now, we have not made any simplifying assumptions about the
prior or weight distributions. To start with, in
Ref.~\cite{Bollweg:2019skg} we have shown that the details of the
prior $p(\omega)$ have no visible effect on the network output. If we
assume a Gaussian prior and, in addition, a Gaussian weight
distribution $q(\omega)$ with the respective means and widths, the
regularization term in Eq.\eqref{eq:loss_bayes1} turns into
\begin{align}
  \kl[q_{\mu,\sigma}(\omega),p_{\mu,\sigma}(\omega)]
  = \frac{\sigma_q^2 - \sigma_p^2 + (\mu_q - \mu_p)^2}{2 \sigma_p^2}
  + \log \frac{\sigma_p}{\sigma_q} \; .
\end{align}
For this form we can use the reparameterization trick to translate an
$\omega$-dependence into a dependence on $\sigma_q$ and
$\mu_q$. Second, we can simplify the loss function by assuming that
the $\omega$-dependent network output in Eq.\eqref{eq:bnn_output} is
described by a Gaussian. This allows us to approximate the likelihood
$p(T|\omega)$ in Eq.\eqref{eq:loss_bayes1} as Gaussian, and the BNN
loss function becomes
\begin{align}
\loss_\text{BNN}
 =& \int d\omega \; q_{\mu,\sigma}(\omega) \; 
   \sum_\text{points $j$} \left[  \frac{\left| \overline{A}_j(\omega) -  A_j^\text{(truth)} \right|^2}{2\sigma_{\text{model},j}(\omega)^2} + \log \sigma_{\text{model},j}(\omega)
   \right] \notag \\
  &+ \frac{\sigma_q^2 - \sigma_p^2 + (\mu_q - \mu_p)^2}{2 \sigma_p^2}
  + \log \frac{\sigma_p}{\sigma_q} \; .
\label{eq:loss_bayes2}
\end{align}
This loss is minimized with respect to the means and standard
deviations of the network weights describing
$q_{\mu,\sigma}(\omega)$. In this setup, the log-likelihood term
includes a trainable uncertainty $\sigma_\text{model}(\omega)$ which
is learned by the network in parallel to the amplitudes. When we
evaluate the likelihood over a mini-batch rather than the full
training dataset, we rescale the normalization of the regularization
term to account for the different numbers of data points.\medskip

The same heteroscedastic loss~\cite{kendall_gal} can be used in
deterministic networks, if we introduce $\sigma_\text{model}$ as a
second trained quantity in addition to the amplitude values. The
Bayesian network setup guarantees that we really capture all
training-related uncertainties correctly, at the expense of splitting
the uncertainty measures $\sigma_\text{model}$ and
$\sigma_\text{pred}$. It also does not make assumptions about a
Gaussian uncertainty of the network output, so we stick to the more
general BNN, even though it might well be possible to use a
deterministic network for similar applications.

\subsubsection*{Network architecture}

We use one Bayesian network trained on the entire training dataset. We
train on amplitudes as a function of phase space with logarithmic
preprocessing,
\begin{align}
  A_j \; \to \; \log \left( 1 + \frac{A_j}{\sigma_A} \right) \; ,
\label{eq:scaling}
\end{align}
where $\sigma_A$ is given by the distribution of the amplitude
values. In addition, all phase space directions are preprocessed by
subtracting the respective mean and dividing by the respective
standard deviation.

The network describing the ($2 \to 3$)-part of the reference process
in Eq.\eqref{eq:proc} consists of four hidden layers with 20 kinematic
input dimensions, $\{ 20,20,30,40 \}$ nodes, and two output dimensions
corresponding to the amplitude and its uncertainty, as illustrated in
Eq.\eqref{eq:bnn_output}. The network has around 6k
parameters. Between the hidden layers we use a $\tanh$ activation
function, while for the last layer we find that a SoftPlus activations
outperforms GeLU slightly and ReLU significantly.  The network is
trained on 90k amplitudes for 400000 epochs with a batch size of 8192
and learning rate of $10^{-4}$, after which we observe no significant
improvement in the loss. We use the Adam optimizer~\cite{adam} with
standard parameters.

\subsubsection*{BNN performance}

\begin{figure}[b!]
  \includegraphics[width=0.48\textwidth, page=2]{c20_e20_performance.pdf}
  \includegraphics[width=0.48\textwidth, page=1]{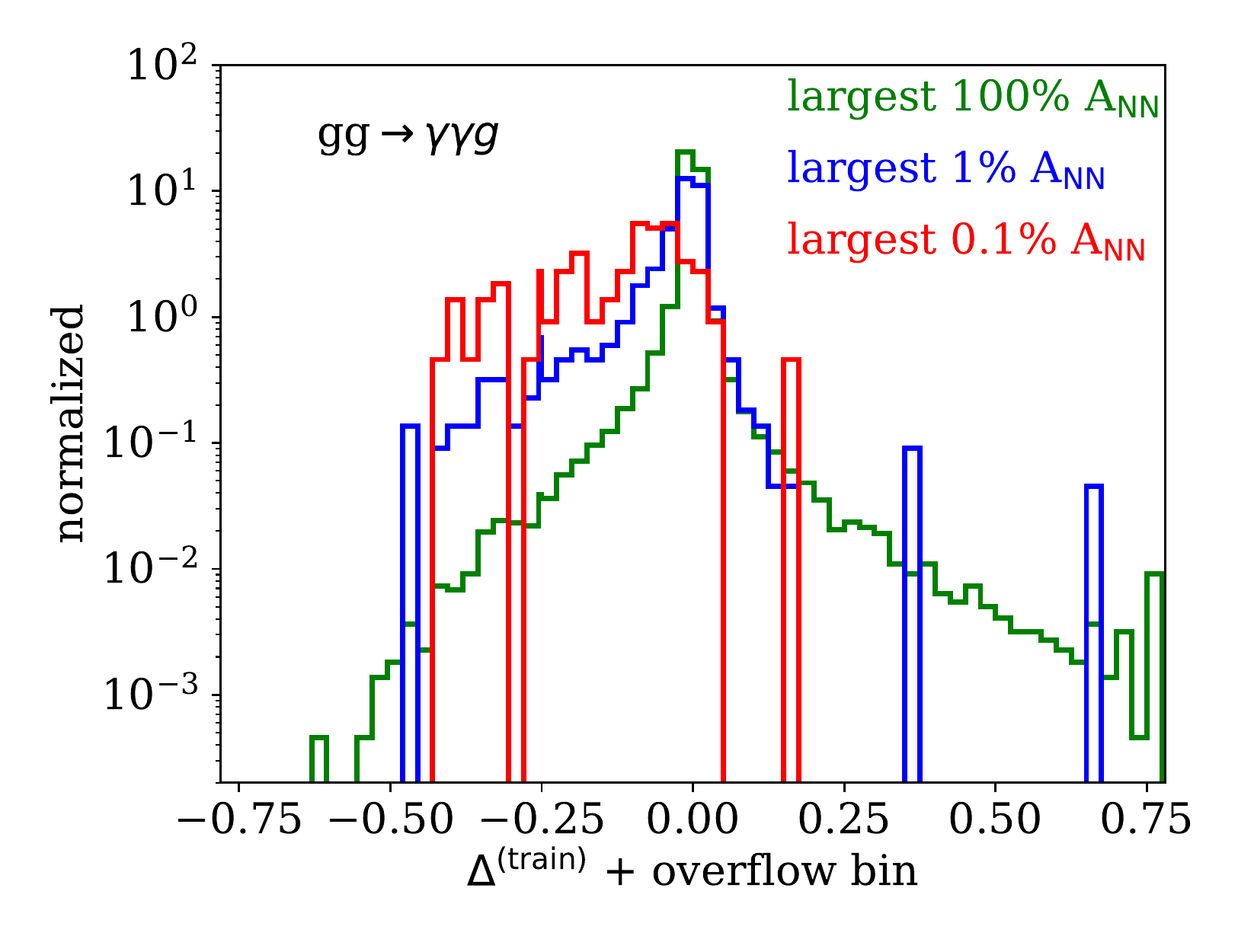} \\
  \includegraphics[width=0.48\textwidth, page=4]{c20_e20_performance.pdf}
  \includegraphics[width=0.48\textwidth, page=3]{c20_e20_performance.pdf}
  \caption{Performance of the BNN in terms of the precision of the
    generated amplitudes, Eq.\eqref{eq:def_deltas}, evaluated on the
    training (upper) and test datasets (lower).}
  \label{fig:bnn_performance}
\end{figure}

As a first test of our BNN, we check the precision with which it
approximates the true amplitudes in the training and test datasets, as
defined in Eq.\eqref{eq:def_deltas}. For
Fig.~\ref{fig:bnn_performance} we split the amplitudes by their
absolute values, to see the effect of the limited training statistics
in the collinear phase space regions. For the complete set of
amplitudes the precision follows an approximate Gaussian with a width
of a few per-mille, for the training and for the test data. This
matches the best available performance from the
literature~\cite{Maitre:2021uaa}, but with a very compact and fast
network.

In the logarithmic panels of Fig.~\ref{fig:bnn_performance} we see
that the tails of the $\Delta$-distributions for the full datasets are
clearly enhanced. The picture changes when we only consider the phase
space points with large amplitudes. For the 0.1\% largest amplitudes
the network is consistently less accurate with a slight tendency of underestimating the amplitudes.
This is the motivation for training a separate network
on the divergent phase space region(s)~\cite{Aylett-Bullock:2021hmo}. As we will see,
the BNN offers an alternative approach which allows the full amplitude to be accurately
described by a single network.\medskip


\begin{figure}[t]
  \includegraphics[width=0.48\textwidth, page=3]{c20_e20_pull}
  \includegraphics[width=0.48\textwidth, page=4]{c20_e20_pull} \\
  \includegraphics[width=0.48\textwidth, page=1]{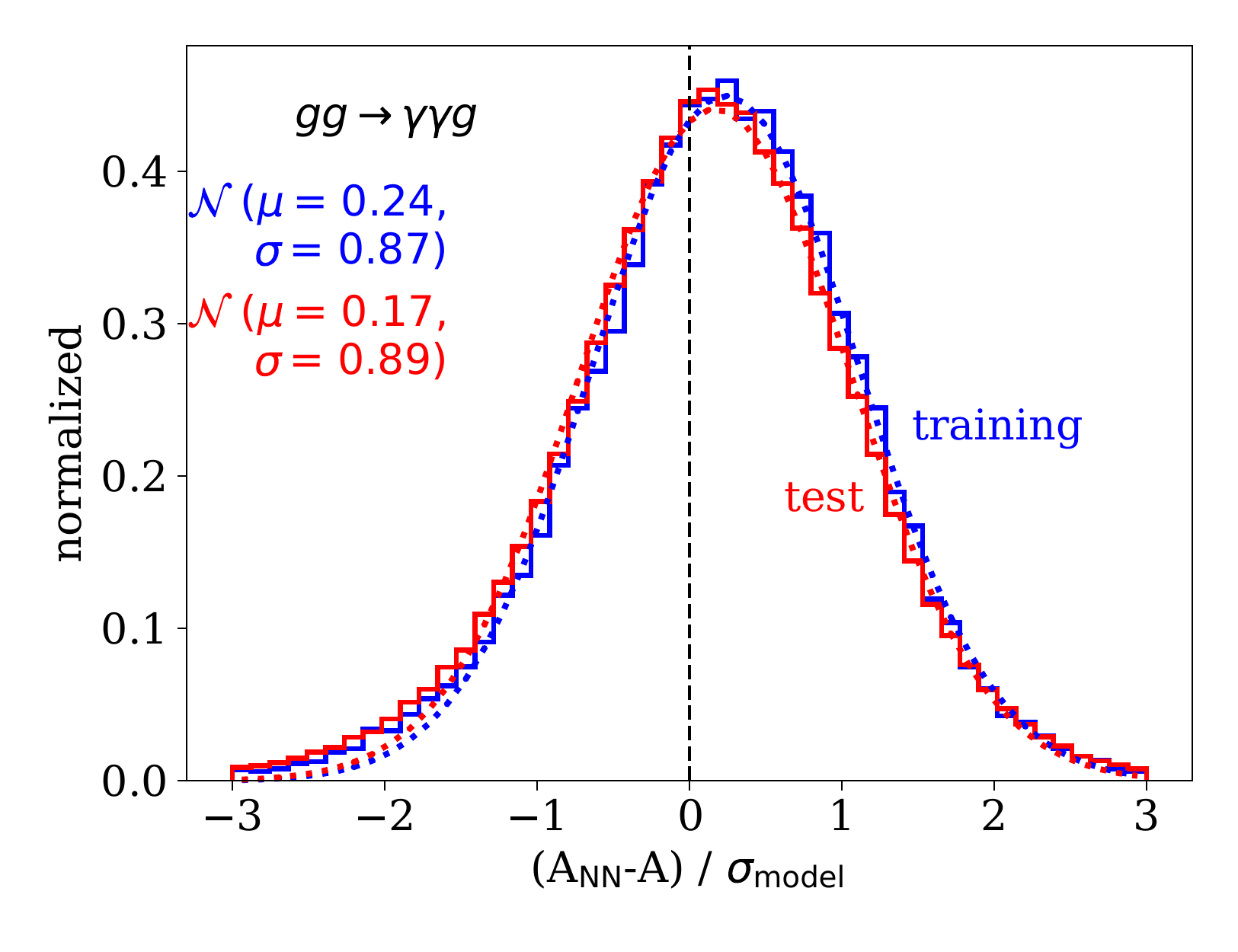}
  \includegraphics[width=0.48\textwidth, page=5]{c20_e20_pull}
  \caption{Pulls for the BNN, defined in Eq.\eqref{eq:pull2} and
    evaluated on the training and test data. The two upper panels show
    the same curve for the weight-dependent pull on a linear and a
    logarithmic axis.}
  \label{fig:bnn_pulls}
\end{figure}

Because the BNN provides us with an uncertainty estimate for the
NN-amplitude, we can define pull variables after integrating over the
weight distributions,
\begin{align}
 t_{\text{model},j} = \frac{ \langle A \rangle_j - A_j^\text{(truth)}}{\sigma_{\text{model},j}}
  \qquad \text{or} \qquad 
 t_{\text{pred},j} = \frac{\langle A \rangle_j - A_j^\text{(truth)}}{\sigma_{\text{pred},j}} \; ,
\label{eq:pull1}
\end{align}
where the point-wise `truth' refers to the training or test datasets
we use to evaluate the pulls. Neither of these pulls have an
$\omega$-dependent counterpart, because their numerators and
denominators are sampled over the network weights independently.  In
the upper panels of Fig.~\ref{fig:bnn_pulls} we see that the two pulls
follow an approximate Gaussian shape, but with a much broader
distribution for the $\sigma_\text{pred}$-based pull because of the
smaller estimated uncertainty. We note that because of the
log-rescaling of Eq.\eqref{eq:scaling} it is not actually the
amplitudes $A$ which should define Gaussian pulls, but their
logarithms. We have explicitly checked that indeed the $\log A$ lead
to a Gaussian, but that given our limited range of relevant amplitudes, the Gaussian shape
translates into an approximately Gaussian shape for the amplitudes
themselves.

Making use of the Gaussian likelihood loss of the BNN,
Eq.\eqref{eq:loss_bayes2}, we can also define the weight-dependent
pull
\begin{align}
t_{\text{model},j}(\omega) = \frac{\overline{A}_j(\omega) - A_j^\text{(truth)}}{\sigma_{\text{model},j}(\omega)} \; .
\label{eq:pull2}
\end{align}
%
%
As part of the loss we can use its distribution as a consistency
condition during network training. Given the Gaussian likelihood loss
we expect a Gaussian distribution of $t_{\text{model},j}(\omega)$,
sampled over $\omega$ according to the Gaussian $q(\omega)$ and over
phase space points $x$. In the upper panels of
Fig.~\ref{fig:bnn_pulls} we see that, again, the pull distribution is
Gaussian in the center, but develops symmetric, enhanced tails roughly
two standard deviations from the mean.\medskip

\begin{figure}[t]
  \centering
  \includegraphics[width=0.48\textwidth]{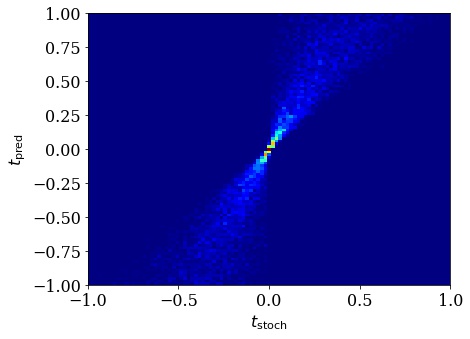}
  \caption{Correlation between the two pulls for the BNN, evaluated on
    the training data.}
  \label{fig:bnn_corr}
\end{figure}

Finally, we need to go back to the definition of the network
uncertainties and understand how the split $\sigma_\text{tot}^2 =
\sigma_\text{model}^2 + \sigma_\text{pred}^2$ can affect improved ways
of training the amplitude network.  We show the two sampled pulls
defined in Eq.\eqref{eq:pull1} in the lower panels of
Fig.~\ref{fig:bnn_pulls}. Both are approximately Gaussian, and the
width of the $t_{\text{model},j}$ distribution is much smaller than
for $t_{\text{pred},j}$. This is an effect of the general observation
that for a well-trained model
\begin{align}
  \sigma_\text{tot} \approx \sigma_\text{model} > \sigma_\text{pred} \; .
\end{align}
The only issue with all pulls shown in Fig.~\ref{fig:bnn_pulls} is
that they come with a slight bias towards positive values, which means
the network slightly overestimates the amplitudes as a whole.  This is
in contrast to the underestimation of the 0.1\% largest
amplitudes observed in Fig.~\ref{fig:bnn_performance}, and we will
target it by improving the network training.

Figure~\ref{fig:bnn_corr} shows very strong correlations between the
two pulls defined in Eq.~\eqref{eq:pull1}. Both pulls correctly
identify the training data points which are not described by the
network well. For our regression tasks with exact amplitudes both
uncertainties are largely induced by the lack of training statistics
especially in the divergent phase space regions, so this correlation
is expected. 

\section{Network boosting} 
\label{sec:boosting}

While the BNN-amplitude results described in the previous section are
promising, the distribution of amplitudes and the pull distributions
indicate potential improvements. We know that for generative networks
we can employ an additional discriminator network to identify poorly
learned phase space regimes~\cite{Butter:2021csz}, the solution is
much simpler for a regression network. In the BNN loss we can compute
the relative deviations between data and network output, or large
pulls, and target these amplitudes directly. Once we control the
network and its uncertainties, we can even think about further
enhancing the training in the direction of an interpolation.

\subsection{Loss-based boosting}
\label{sec:boosting_loss}

\begin{figure}[t]
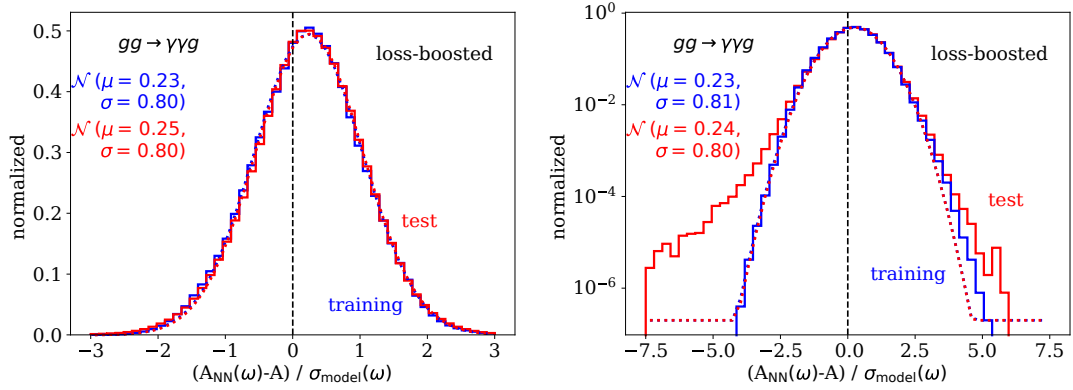

  \includegraphics[page=3, width=0.48\textwidth]{loss_feedback_pull.pdf}
  \includegraphics[page=4, width=0.48\textwidth]{loss_feedback_pull.pdf} 
  \caption{Pulls for the loss-boosted BNN, defined in
    Eq.\eqref{eq:pull2} and evaluated on the training and test
    data. The two panels show the same results on a linear and a
    logarithmic axis. All curves can be compared to the BNN results
    without boosting in Fig.~\ref{fig:bnn_pulls}.}
  \label{fig:bbnn_pulls}
\end{figure}

Because the BNN loss in Eq.\eqref{eq:loss_bayes2} represents a
Gaussian log-likelihood, we can modify it and require a higher
precision for those phase space points which according to the BNN
uncertainty are not yet learned well. In practice, this is equivalent
to feeding these training data points $n_j$ times into the 
computation of the BNN loss
\begin{align}
\loss_\text{Boosted BNN}
 =& \int d\omega \; q_{\mu,\sigma}(\omega) \; 
   \sum_\text{points $j$} n_j \times \left[  \frac{\left| \overline{A}_j(\omega) -  A_j^\text{(truth)} \right|^2}{2\sigma_{\text{model},j}(\omega)^2} + \log \sigma_{\text{model},j}(\omega)
   \right] \notag \\
  &+ \frac{\sigma_q^2 - \sigma_p^2 + (\mu_q - \mu_p)^2}{2 \sigma_p^2}
  + \log \frac{\sigma_p}{\sigma_q} \; .
\label{eq:loss_bayes3}
\end{align}
As mentioned for Eq.\eqref{eq:loss_bayes2}, the regularization has to
be adjusted for the additional amplitudes in the boosted training
sample.  This feedback training is similar to simple boosting
algorithms for decision trees, where amplitudes for which the decision
tree gives a wrong result are enhanced with additional weights. In our
simple approach we duplicate some training amplitudes, or equivalently
increases their weights in discrete steps.

In a first stage, we improve the self-consistency of the network with
the initial assumptions and boost the network training for amplitudes
with a large $\omega$-dependent pull, Eq.\eqref{eq:pull2}.  
In five iterations we identify the amplitudes with values of
$t_{\text{model},j}(\omega)$, which are more than two standard
deviations away from the mean and increase their contribution to the
loss function in Eq.\eqref{eq:loss_bayes3} by values $n_j = 5$. This
is done four times. After adding the weights we continue the training
on the enlarged datasets. For the next training cycle we again add
weights to the amplitudes which now have pulls more than two standard
deviations away.  Each training ends when we see no more significant
change to the loss which usually takes around 2000 epochs. This
boosting forces the network towards a more self-consistent description
of the tails of the pull distributions. We checked that small
variations of $n_j$ or the number of cycles do not have a significant
impact on these improvements.

\begin{figure}[t]
  \includegraphics[page=2, width=0.48\textwidth]{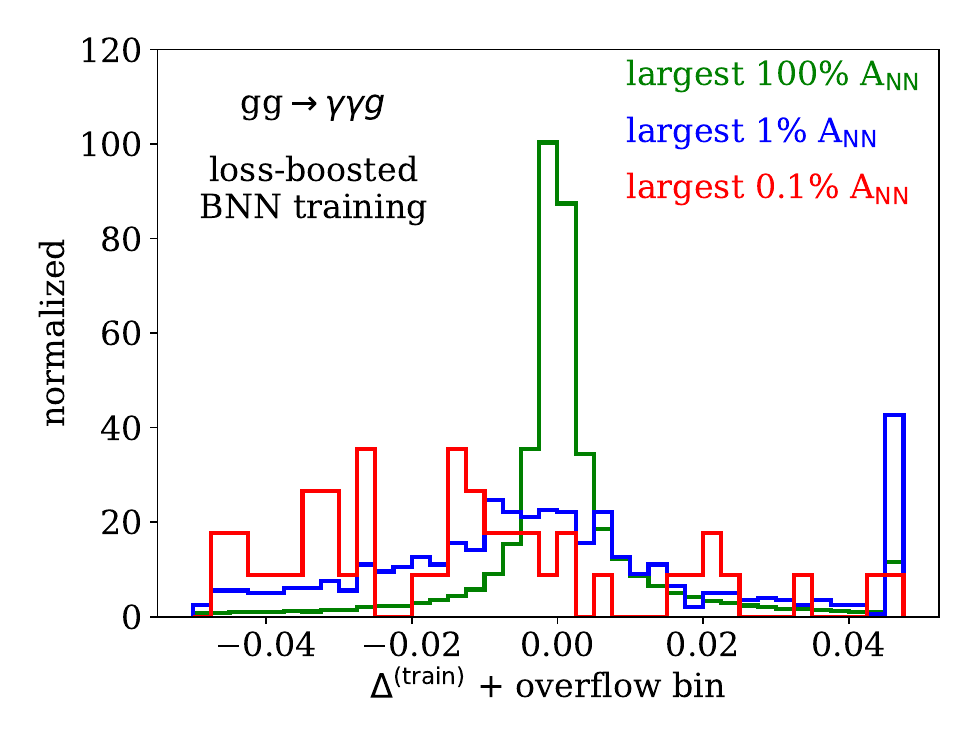}
  \includegraphics[page=4, width=0.48\textwidth]{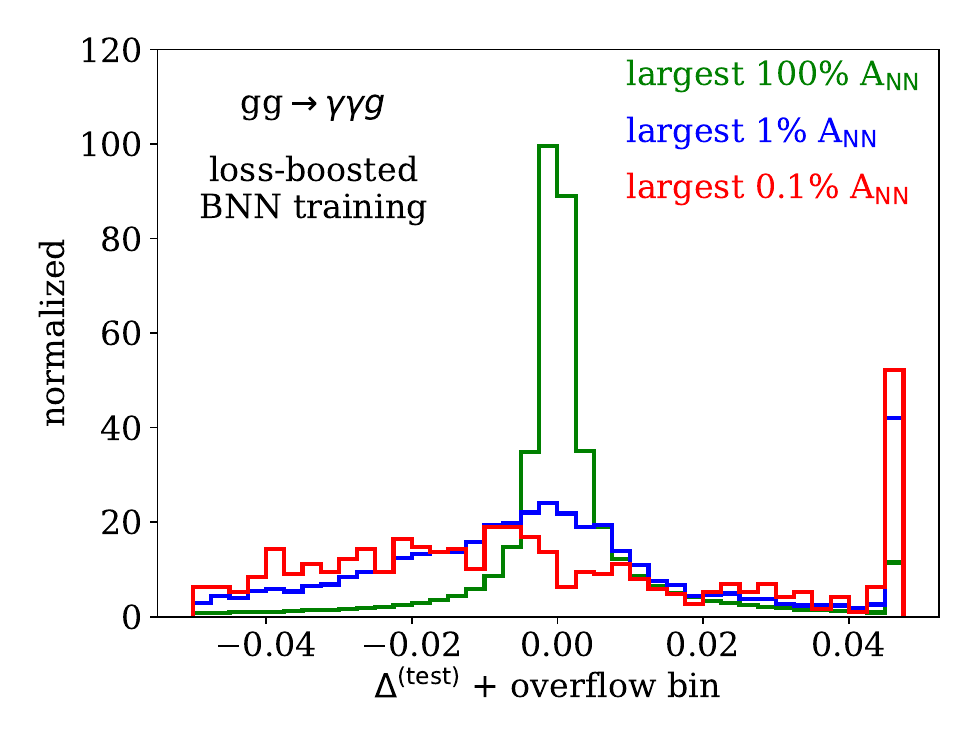}
  \caption{Performance of the loss-boosted BNN in terms of the
    precision of the generated amplitudes, Eq.\eqref{eq:def_deltas},
    evaluated on the training and test datasets on a linear (left) and
    logarithmic (right) axis. The curves can be compared to the BNN
    results without boosting in Fig.~\ref{fig:bnn_performance}.}
  \label{fig:bbnn_performance}
\end{figure}

In Fig.~\ref{fig:bbnn_pulls} we show the pulls from the boosted
Bayesian neural network, boosted based on the self-consistency
of the loss measured by the pull.  We see a significant improvement
for $t_\text{model}(\omega)$, the parameter we target with our
boosting. One would naively expect the corresponding distribution to
assume a Gaussian shape with unit width. However, first of all our
loss-based boosting only moves amplitudes from the tails into the
bulk, which means that the tails of the boosted pull distributions
should be low. Second, the pulls entering the loss and the pulls shown
in Fig.~\ref{fig:bbnn_pulls} are different because the loss includes
weights for high-pull amplitudes. In combination, both effects explain
the narrower Gaussian for $t_\text{model}(\omega)$. In the logarithmic
version we also see a visible over-training though loss-boosting.

%

Moving on to the precision for the amplitudes, we see in
Fig.~\ref{fig:bbnn_performance} that the loss-boosting only has a mild 
impact on the $\Delta$-distributions. It does not significantly
improve the precision of the amplitudes compared to
Fig.~\ref{fig:bnn_performance}, so we need a second boosting step.


\subsection{Performance boosting}
\label{sec:boosting_proc}

\begin{figure}[b!]
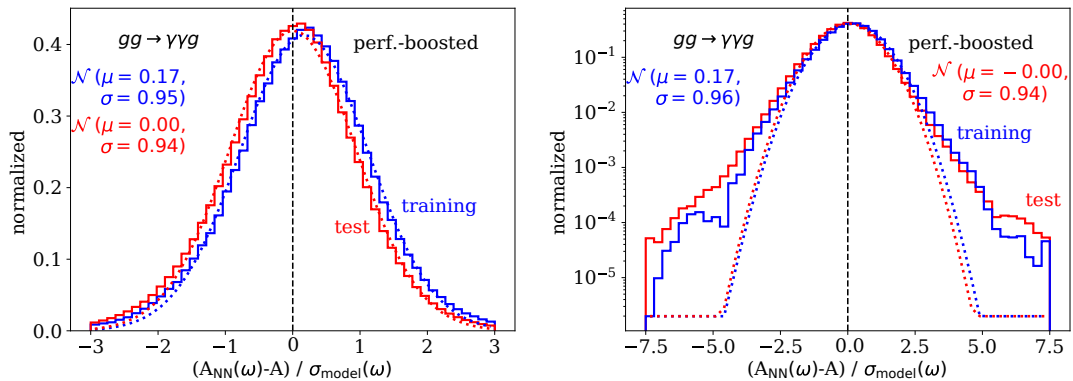

  \includegraphics[page=3, width=0.48\textwidth]{process_feedback_pull.pdf}
  \includegraphics[page=4, width=0.48\textwidth]{process_feedback_pull.pdf} 
  \caption{Pulls for the performance-boosted BNN, defined in
    Eq.\eqref{eq:pull2} and evaluated on the training and test
    data. The two panels show the same results on a linear and a
    logarithmic axis. All curves can be compared to the BNN results
    without boosting in Fig.~\ref{fig:bnn_pulls} and the loss-boosted
    results in Fig.~\ref{fig:bbnn_pulls}.}
  \label{fig:bbnn_pulls2}
\end{figure}

Given that the loss-boosting in the previous section worked for the
uncertainty estimate but only had a modest effect on the performance
of our amplitude network, we proceed to a more powerful boosting
strategy. Independent of the self-consistency of the network, we know
at the training level which amplitudes challenge the network. This
means we can select them with the goal of improving the training for
the largest amplitudes.  The difference between a general loss
boosting and this process-dependent strategy is that now we target the
largest and most poorly learned amplitudes by selecting them based on
$\sigma_\text{tot}$. We choose the 200 amplitudes with the largest
uncertainty $\sigma_\text{tot}$ and add three additional copies to
the training dataset. This process is repeated 20 times, where each
training ends when we see no more significant change to the loss which is usually around 2000 epochs.

In Fig.~\ref{fig:bbnn_pulls2} we first see that the process-specific
performance boosting broadens the pull distributions and this way reverses some of
the beneficial effects of the loss-boosting on
$t_\text{model}(\omega)$. However, the widths of the boost
distributions remains below one, and the bias towards larger
amplitudes is removed. This is true for the training data and for the
test data. In addition, the consistency with the Gaussian shape is
broken symmetrically for too small and too large amplitudes, again
consistently for training and test data.  Given that the two boostings
target different amplitudes and effectively compete with each other,
this pattern is expected.

\begin{figure}[t]
  \includegraphics[page=2, width=0.48\textwidth]{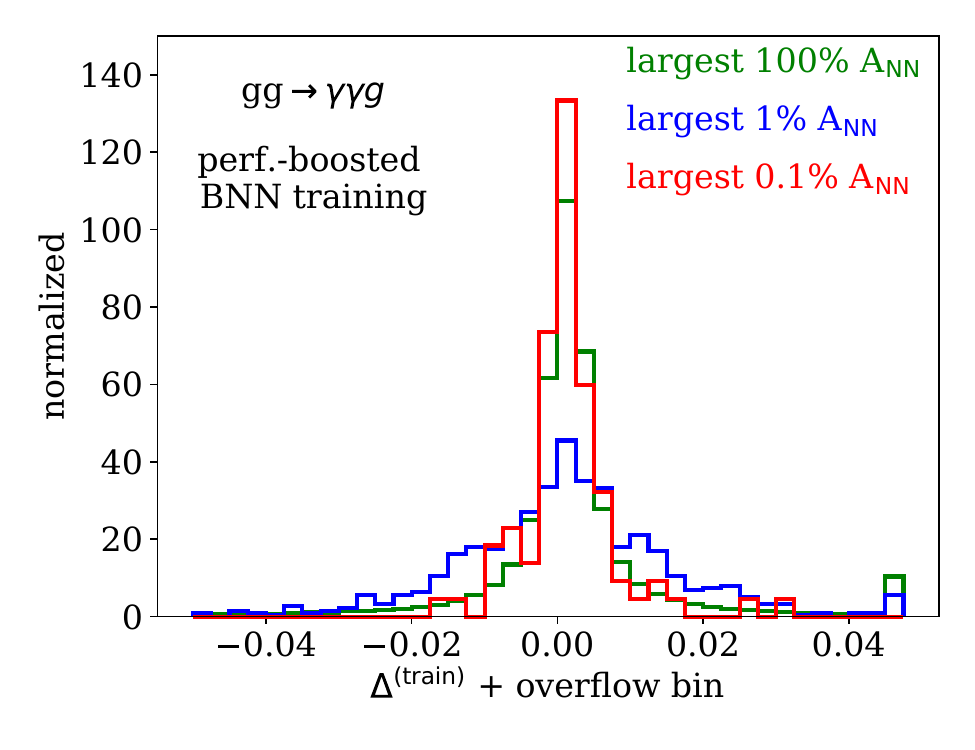}
  \includegraphics[page=4, width=0.48\textwidth]{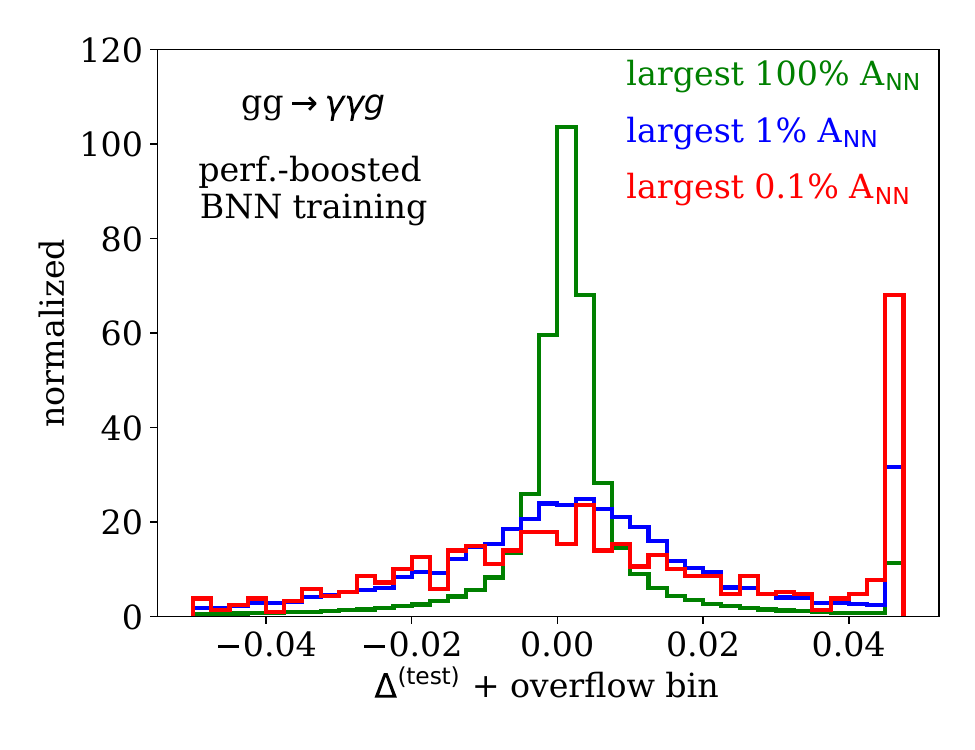}
  \caption{Performance of the performance-boosted BNN in terms of the
    precision of the generated amplitudes, Eq.\eqref{eq:def_deltas},
    evaluated on the training (left) and test datasets (right). All
    curves can be compared to the BNN results without boosting in
    Fig.~\ref{fig:bnn_performance} and the loss-boosted results in
    Fig.~\ref{fig:bbnn_performance}.}
  \label{fig:bbnn_performance2}
\end{figure}

The positive impact on the large amplitudes can be seen more clearly
in Fig.~\ref{fig:bbnn_performance2}. Evaluated on the training data,
the 0.1\% largest amplitudes now show a clear peak at small
$\Delta^\text{train}$, consistent with all other amplitudes. This
means the network has learned all amplitudes in the training dataset
equally well.  This effect translates to the test sample
qualitatively, so the performance on the test data improves after
performance-boosting, but this improvement is less pronounced than for the
training data. This means that, at the expense of an overtraining, we
have improved our network from a fit-like description to an
interpolation-like description of the largest amplitudes.

The pattern observed by performance-boosting points to a conceptual
weakness of standard network training when it comes to precision
applications. If we stop the network training at the point where the
performance on a training sample exceeds the performance on the test
sample, we miss the opportunity of improving the network on the test
and training data, but at a different rate. Overtraining is, per se,
not a problem, as we know from applications of interpolation to
describe data. The only challenge from such a network overtraining is
a reliable uncertainty estimate from the generalization, for which we
propose an appropriate scheduling of loss-boosting and
performance-boosting.

\subsection{Effect of training statistics}

\begin{figure}[t]
    \includegraphics[width=0.48\textwidth, page=1]{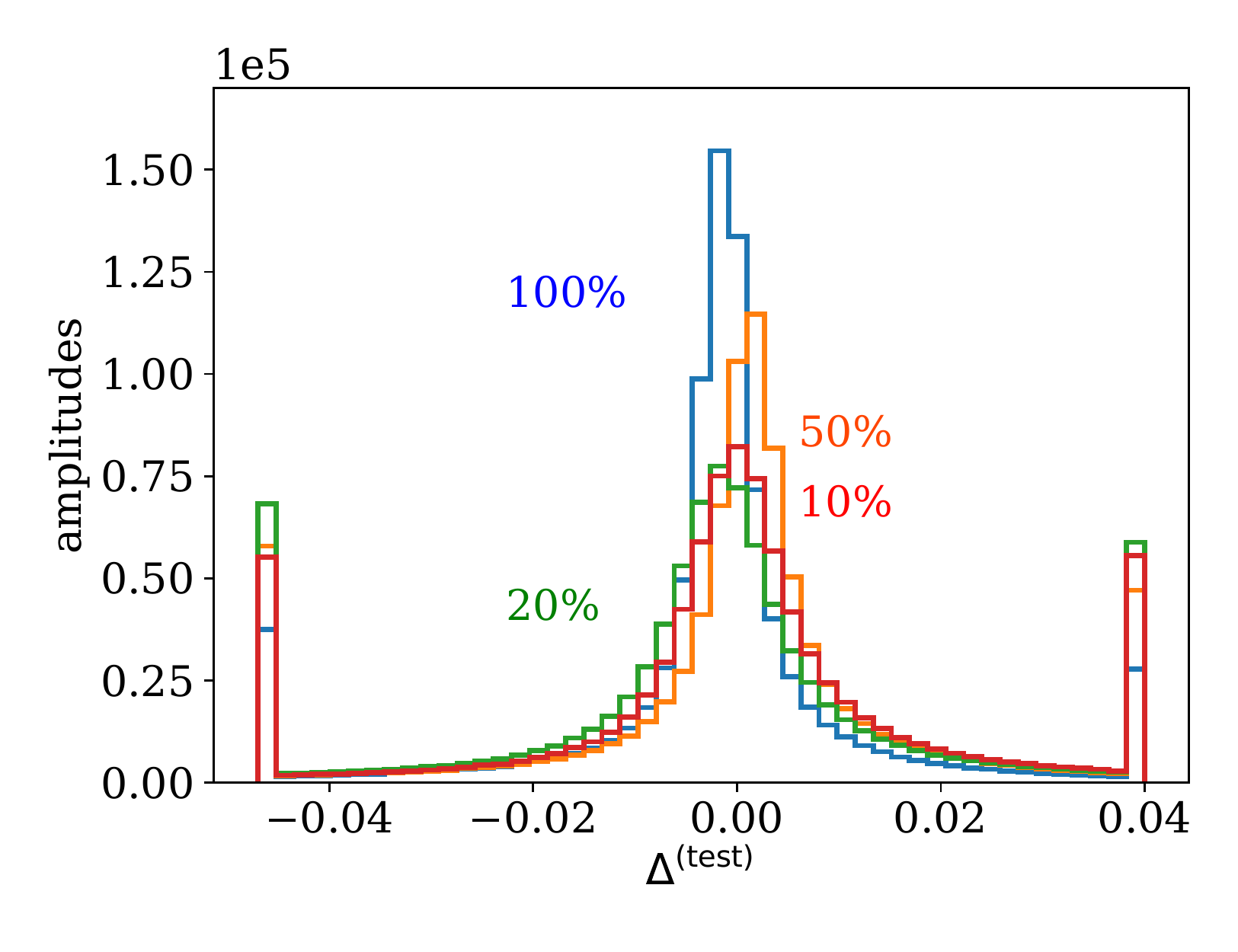}
    \includegraphics[width=0.48\textwidth, page=4]{reduced_data.pdf}
    \caption{Performance of the BBN for all amplitudes (left) and a
      performance-boosted BNN for the largest 1\% of all amplitudes
      (right), after training on different fractions of the full
      training dataset.}
    \label{fig:less_data}
\end{figure}

Given that our amplitude-BNN has successfully learned the amplitudes
for the partonic process $gg \to \gamma \gamma g$ well below the
percent level, with a small and simple network and 90k training
points, we can ask the question how much training data we actually
need for a precision amplitude network. For this study we use the same
BNN as before, including loss-boosting and performance-boosting, but
trained on a reduced dataset of
\begin{align}
  10\% \quad (9.000 \; \text{amplitudes})
  \quad \cdots \quad
  100\% \quad (90.000 \; \text{amplitudes}) \; .
\end{align}
In Fig.~\ref{fig:less_data} we show the corresponding
$\Delta$-distributions for the test dataset.  Our smallest training
dataset contains 9000 amplitudes, which turn out sufficient to train
our network with its 6192 parameters. The corresponding network
reproduces the test data well, albeit with sizeable overflow
bins. Increasing the amount of training data improves the precision of
the network, but relatively slowly. We observe the same level of
improvement for all amplitudes and for the 1\% largest amplitudes. For
the latter we only show results after process boosting, without any
boosting the quality of the low-statistics training is comparably
poor.

\section{Kinematic distributions}

After illustrating the performance of the amplitude network in a
somewhat abstract manner, we can also show 1-dimensional kinematic
distributions. The integration of the remaining phase space dimension
requires a little care, because we cannot just integrate the
uncertainties together with the central values for the amplitudes.

For the central values we combine the amplitudes with phase space
sampling. For example applying the simple
\textsc{Rambo}~\cite{Kleiss:1985gy} algorithm we identify the phase
space weights with $A$. A 1-dimensional distribution is generated
through bins which collect the sum of the amplitudes in the remaining
phase space directions.  The histogram value for a bin $k$
is 
%
\begin{align}
  h_k
  = \sum_{j=1}^N  A_j \; .
\end{align}
%
To use the amplitudes predicted by the BNN we have to add the sampling
over the weights $\omega$.  By replacing the truth amplitudes with the
NN-amplitudes we can target the uncertainties from the modelling of the
amplitudes through the BNN.  In analogy to Eq.\eqref{eq:expectations}
and omitting the index $k$ for the histogram we first extract a
central histogram value as
\begin{align}
  \langle h \rangle
  &= \int d\omega \;  q(\omega) \, \sum_j \overline{A}_i(\omega) \notag \\
  &= \int d\omega \;  q(\omega) \,  \overline{h} (\omega) 
  \qquad \text{with} \qquad
  \overline{h} (\omega)
  = \sum_j  \bar{A}_j(\omega)  \; .
\end{align}
Again in analogy to Eqs.\eqref{eq:sig_pred} and~\eqref{eq:sig_model}
we define the absolute uncertainties on the bin entry as
\begin{align}
  \sigma_{h,\text{pred}}^2
  &= \int d\omega \; q(\omega) \; \Big[ \overline{h} (\omega) - \langle h \rangle  \Big]^2
  \notag \\
  \sigma_{h,\text{model}}^2
  &= \int d\omega \; q(\omega) \; \Big[ \overline{h^2} (\omega) - \overline{h} (\omega)^2 \Big] \; .
\label{eq:h_model}
\end{align}
The total uncertainty is again $\sigma_{h,\text{tot}}^2 =
\sigma_{h,\text{model}}^2 + \sigma_{h,\text{pred}}^2$.  We can
simplify $\sigma_{h,\text{model}}$ further. In all of the above
formulas $h$ is just a sum over amplitudes. If we assume that the
corresponding $\sigma_\text{model}$ values are uncorrelated, we can
relate $\sigma_{h,\text{model}}$ to $\sigma_\text{model}$ by
exchanging the sum and the variance,
\begin{align}
  \sigma_{h,\text{model}}^2
&= \int d\omega \; q(\omega) \, \text{Var} (\overline{h}(\omega)) \notag \\
&= \int d\omega \; q(\omega) \, \text{Var}\left( \sum_j A_j(\omega) \right) 
= \int d\omega \; q(\omega) \sum_j \text{Var} \left( A_j(\omega) \right) \notag \\
&=\int d\omega \; q(\omega)  \sum_j \sigma_{\text{model}, j}^2(\omega) 
  \equiv \XXLangle   \sum_j \sigma_{\text{model}, j}^2(\omega)  \XXRangle
 \; .
 \label{eq:hist_model_derivation2}
\end{align}
\medskip

While we assume uncorrelated uncertainties for $\sigma _\text{model}$
we cannot do the same for $\sigma _\text{pred}$.  To compute $\sigma
_\text{pred}$ we first sample a set of weight configurations, which
turns our BNN into an ensemble of neural networks, and then use each
of these neural networks to compute the corresponding histogram
value. Computing the standard deviation of these values gives us an
estimate for $\sigma_{h,\text{pred}}$.  By sampling from the weight
distributions we change the neural network itself and all of its
predictions.  To assume that these changes are uncorrelated for
different amplitudes seems not exceptionally well justified.\medskip

\begin{figure}[t]
  \includegraphics[page=4, width=0.495\textwidth]{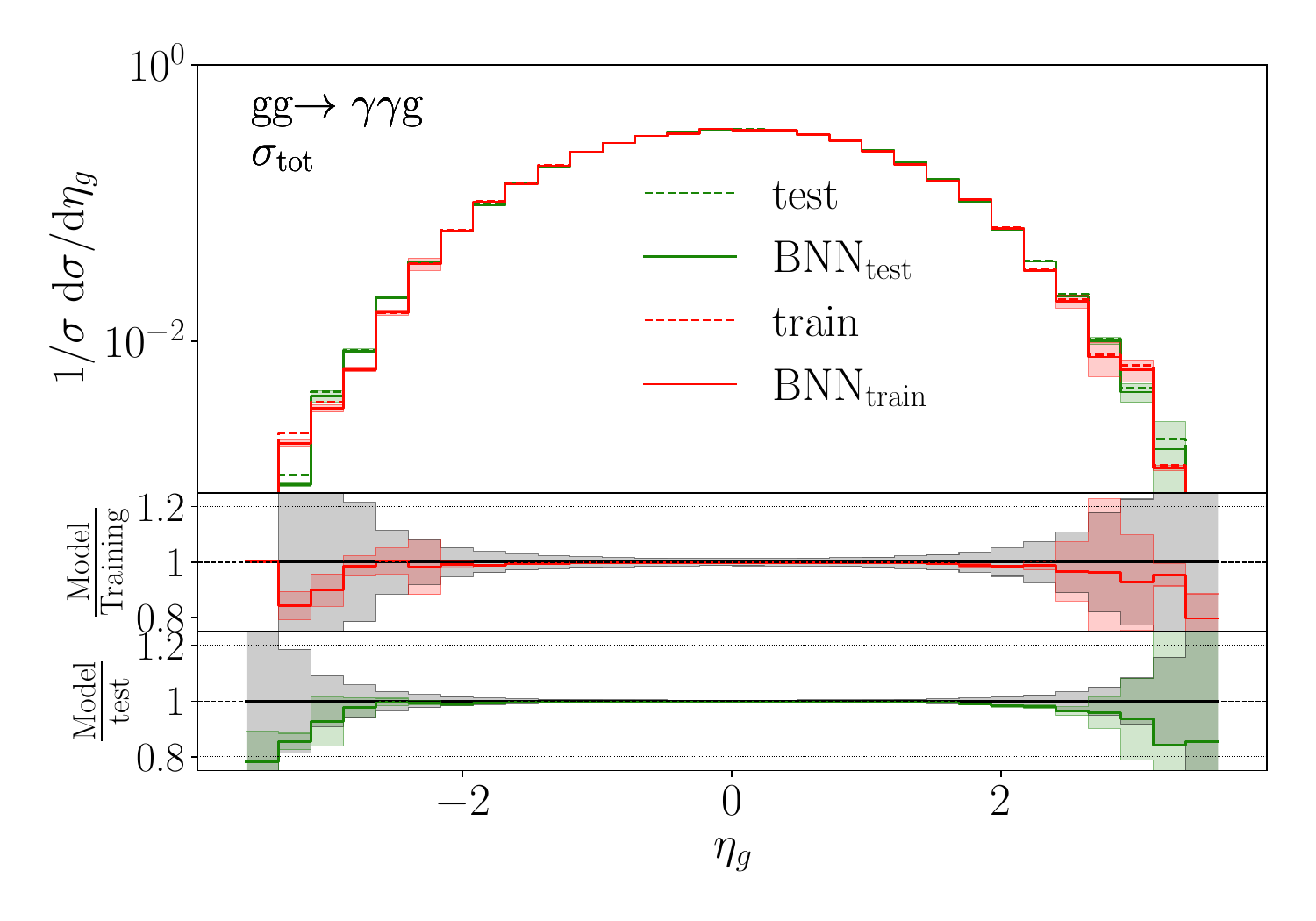}
  \includegraphics[page=12, width=0.495\textwidth]{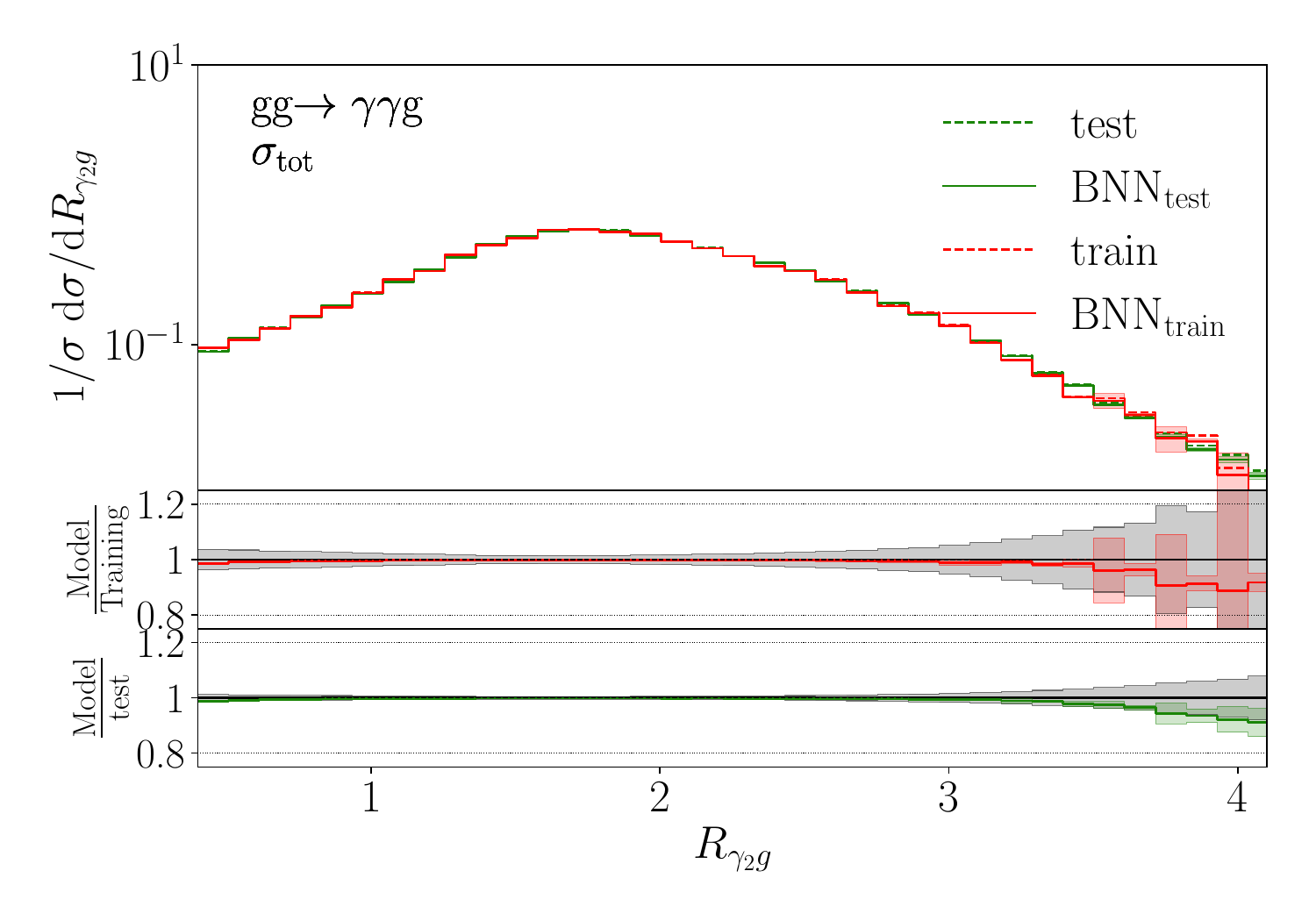}
  \includegraphics[page=4, width=0.495\textwidth]{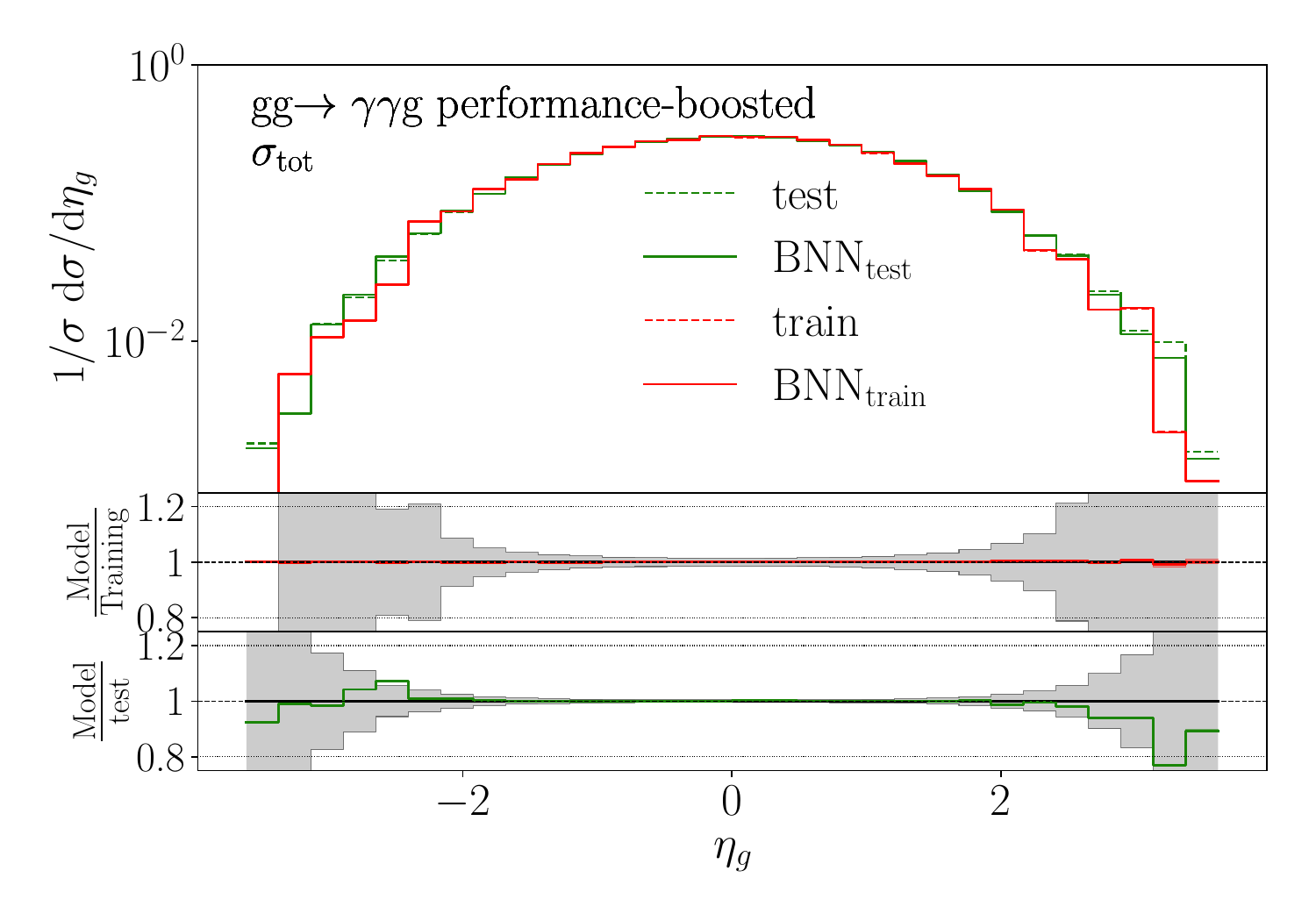}
  \includegraphics[page=12, width=0.495\textwidth]{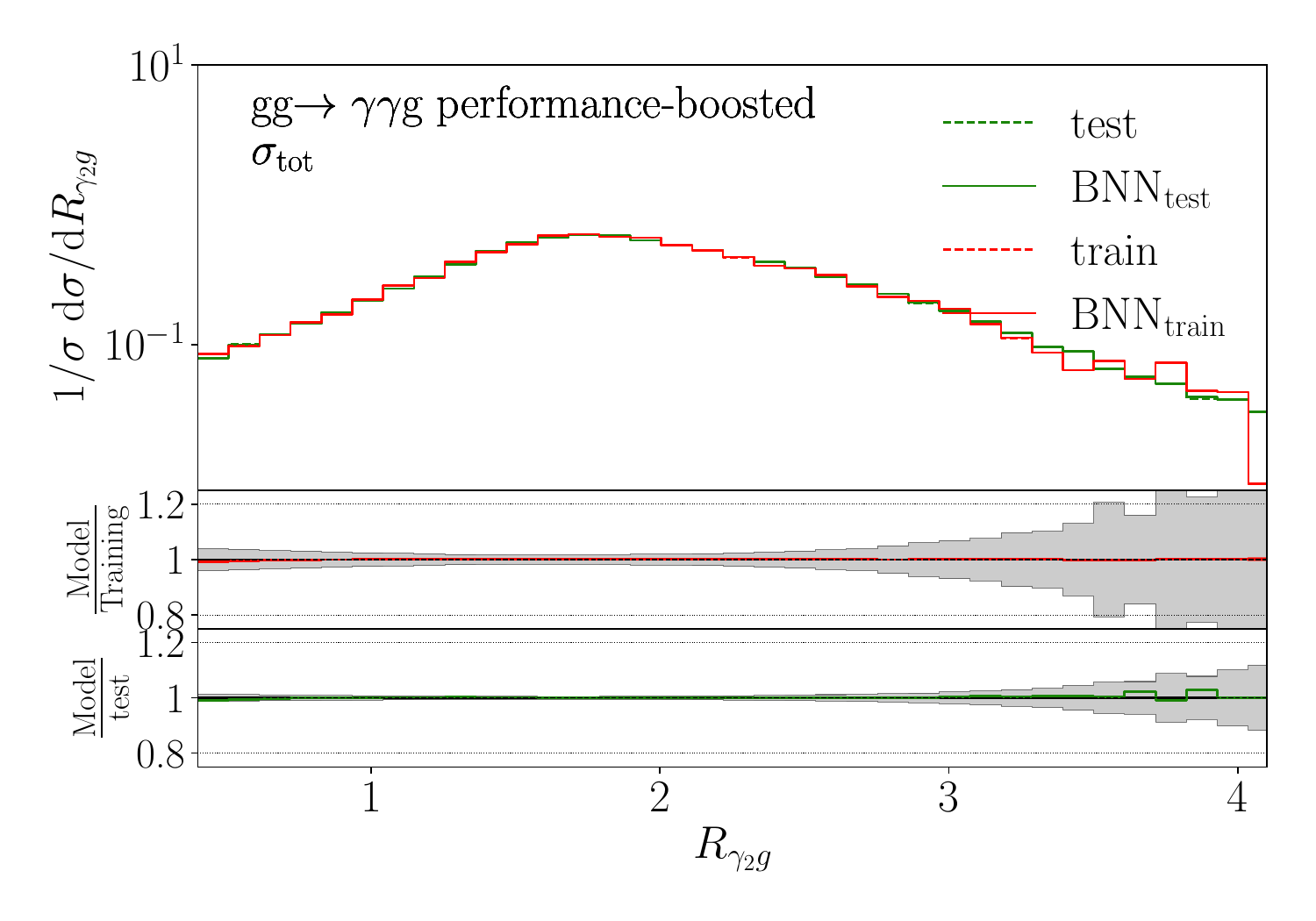}
  \caption{Kinematic distributions from the BNN without boosting
    (upper) and after performance-boosting (lower). The grey error bars in
    the lower panels indicate the statistical limitation of the
    training and test data.}
  \label{fig:distris}
\end{figure}

Based on this procedure we show BNN-amplitude results for a set of
kinematic distributions in the upper panels of
Fig.~\ref{fig:distris}. We see the effect of limited training data
towards the end of the different kinematic distributions, where the
agreement between the NN-amplitudes and truth deteriorates. For our
reference process this happens for $|\eta_g| \gtrsim 2.5$ or
$|\eta_\gamma| \gtrsim 1.5$. Still, the BNN uncertainty estimate
covers the deviation from the truth reliably.

In the lower panels of Fig.~\ref{fig:distris} we see that after
performance-boosting the BNN predictions agree with the training data
spectacularly well. This is the goal of the boosting and leads to the
network learning all features in the training data extremely well. In
the phase space regions where the regular BNN precision is limited by
sparse and large training amplitudes, the improved agreement between
NN-amplitudes and the training data carries over to the test data at a
level that the network prediction is significantly improved. The
uncertainties for the training data still cover the deviations from
the truth, but unlike the central values this uncertainty estimate
does not generalize correctly to the test data. This structural issue
with process boosting could be ameliorated by alternating between
loss-boosting and performance-boosting, until the specific requirements of
a given analysis in precision and uncertainty estimates are met.

\section{Outlook}

Learning loop-amplitudes for LHC simulations is a classic ML-problem,
because we need to train a precision network only once to provide a
much faster simulation tool which can be used many times. In this
application neural networks really work like better fits to the
training data. Unlike for many other network applications, the
training amplitudes are not noisy, which means we would like to
reproduce the training amplitudes exactly, supplemented with a
controlled uncertainty over all of phase space. To provide a reliable
uncertainty map over phase space, we can rely on Bayesian regression
networks~\cite{Kasieczka:2020vlh}.

The precision task reminds us of an interpolation rather than a fit,
which means we need modify our ML-approach conceptually. If we are
willing to accept a certain amount of overtraining, we can
significantly improve the network training through boosting certain
amplitudes. Because the Bayesian network provides a reliable
uncertainty estimate, we can select the to-be-boosted amplitudes based
on their deviation from the training data in units of the
uncertainty. This loss-based boosting simply improves the
self-consistency of the Bayesian network training. In a second step,
we can boost training amplitudes just based on their absolute
uncertainty. This selection helps with the precision for a given
process, and because we use the absolute uncertainty we typically
focus on the largest amplitude values.

We have applied Bayesian network training and the two strategies of
amplitude boosting to the partonic process $gg \to \gamma \gamma
g$~\cite{Aylett-Bullock:2021hmo}. We have first found that the network
amplitudes agree with the true amplitudes at the sub-percent level,
for the training data and for a test dataset. For the 1\% largest
amplitudes an agreement at the percent level required process-specific
performance boosting. For 1-dimensional kinematic distributions we
have seen that the performance-boosting allows for extremely precise
predictions in kinematic tails, albeit with a somewhat reduced
performance in the uncertainty estimate for the test dataset. This can
be improved by alternating between process and loss boosting in order
to retain improved uncertainty estimation and increased performance
which will be subject of future studies.

Finally, we have checked what happens with our boosted Bayesian
network training when we reduce the number of training amplitudes from
90k to 9k and found that thanks to the boosteing this only leads to a
mild decrease in the network precision. This leaves us confident that
boosted amplitude training with its shift from a fit-like to
interpolation-like objective provides us with highly efficient
surrogate models whenever the generation of training data is
CPU-intensive.

\section*{Acknowledgments}

First, we would like to thank Manuel Hau{ss}mann for introducing us to
Bayesian networks.  We thank Steffen Schumann for many enlightening
discussions and Frank Krauss for regularizing any unreasonable
enthusiasm. We are also grateful to Joseph Aylett-Bullock and Ryan
Moodie for helpful conversations. This research is supported by the
Deutsche Forschungsgemeinschaft (DFG, German Research Foundation)
under grant 396021762 -- TRR~257: \textsl{Particle Physics
  Phenomenology after the Higgs Discovery}, through Germany's
Excellence Strategy EXC 2181/1 - 390900948 (the Heidelberg STRUCTURES
Excellence Cluster) and the European Union's Horizon 2020 research and
innovation programmes \textit{High precision multi-jet dynamics at the
  LHC} (consolidator grant agreement No 772009).

\clearpage
\appendix
\section{$(2 \to 4)$-process}

We know that that increasing the number of particles in the final
state leads to a significant drop in network
performance~\cite{Badger:2020uow,Maitre:2021uaa,Aylett-Bullock:2021hmo,Bothmann:2020ywa,Gao:2020zvv}. We
illustrate how we can use the boosted BNN training to improve network
predictions for the $(2 \to 4)$-process
\begin{align}
gg \to \gamma \gamma gg \; .
\label{eq:proc}
\end{align}
For this process we use a network with seven hidden layers, 24
kinematic input dimensions, $\{ 32,64,256,512,128,64,32 \}$ nodes, and
two output dimensions corresponding to the amplitude and its
uncertainty. This larger network has around 600k parameters.  The training dataset contains around 90k amplitudes. Aside from these
changes, we apply the same basic BNN training with two levels of loss
boosting and process-specific performance boosting.

\begin{figure}[b!]
  \includegraphics[width=0.495\textwidth, page=2]{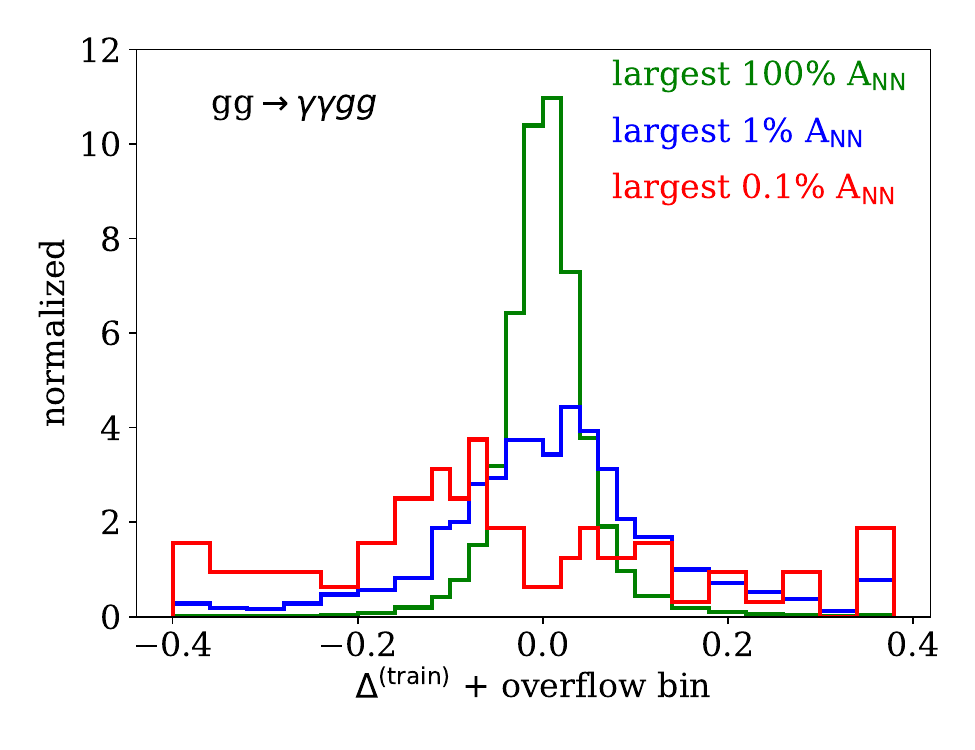}
  \includegraphics[width=0.495\textwidth, page=4]{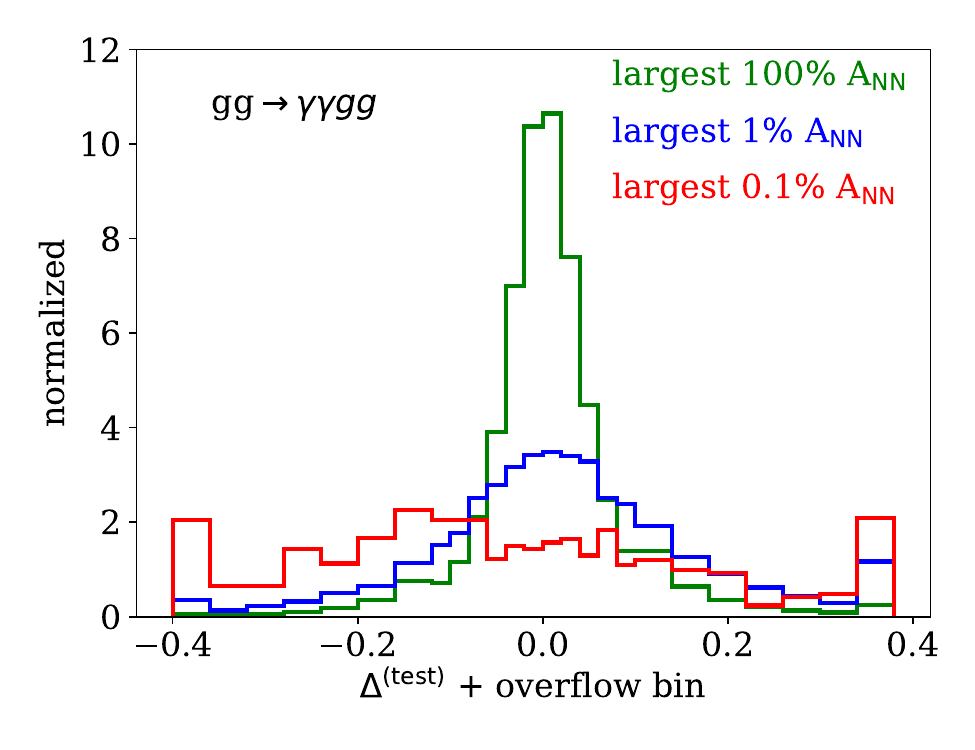}
  \includegraphics[width=0.495\textwidth, page=2]{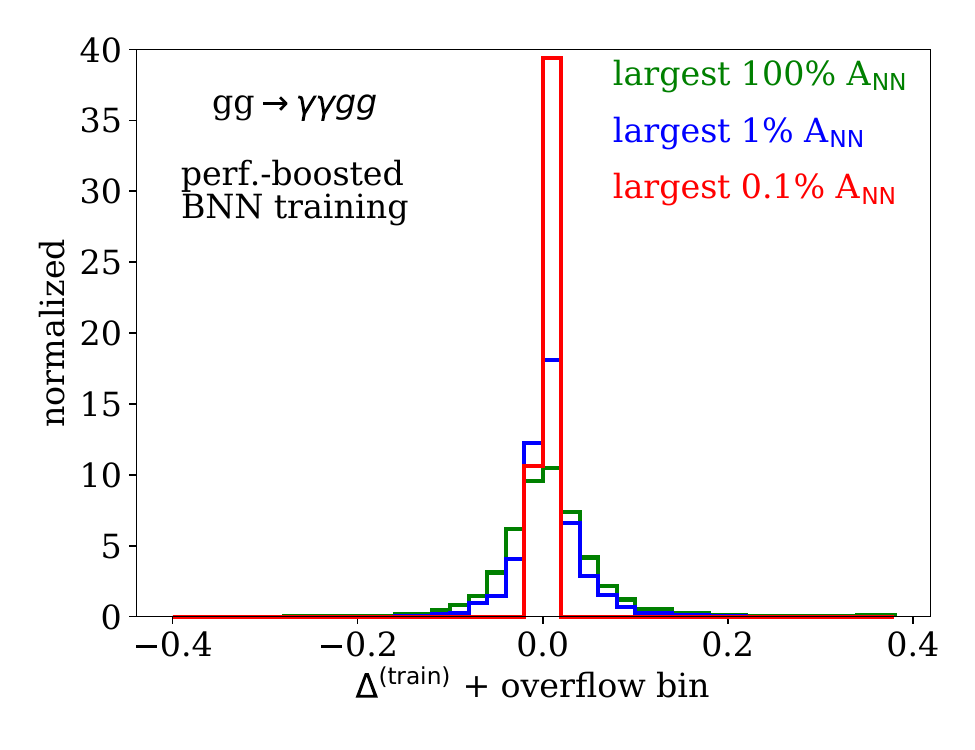}
  \includegraphics[width=0.495\textwidth, page=4]{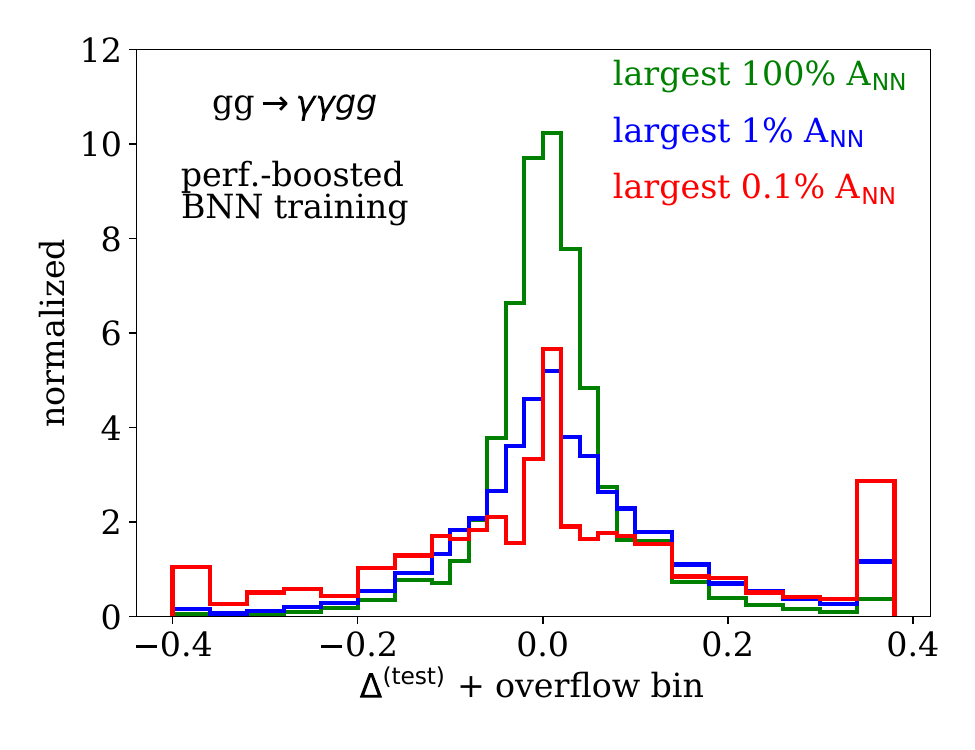}
  \caption{Performance of the basic (upper) and performance-boosted
    (lower) BNNs for the $(2 \to 4)$-process in terms of the precision
    of the generated amplitudes, Eq.\eqref{eq:def_deltas}, evaluated
    on the training (left) and test datasets (right).}
  \label{fig:2to4perf}
\end{figure}

As for the $(2 \to 3)$-process we first show the performance of the
network training in Fig.~\ref{fig:2to4perf}. While the overall scale
of the agreement has increased for the sub-percent level to the
percent level, we still see that the network has learned the largest
amplitudes extremely well after process boosting. Unlike for the
standard BNN, there is a certain amount of overtraining after
performance-boosting, indicating the shift from a fit-like network
training to an interpolation-like training.

\begin{figure}[t]
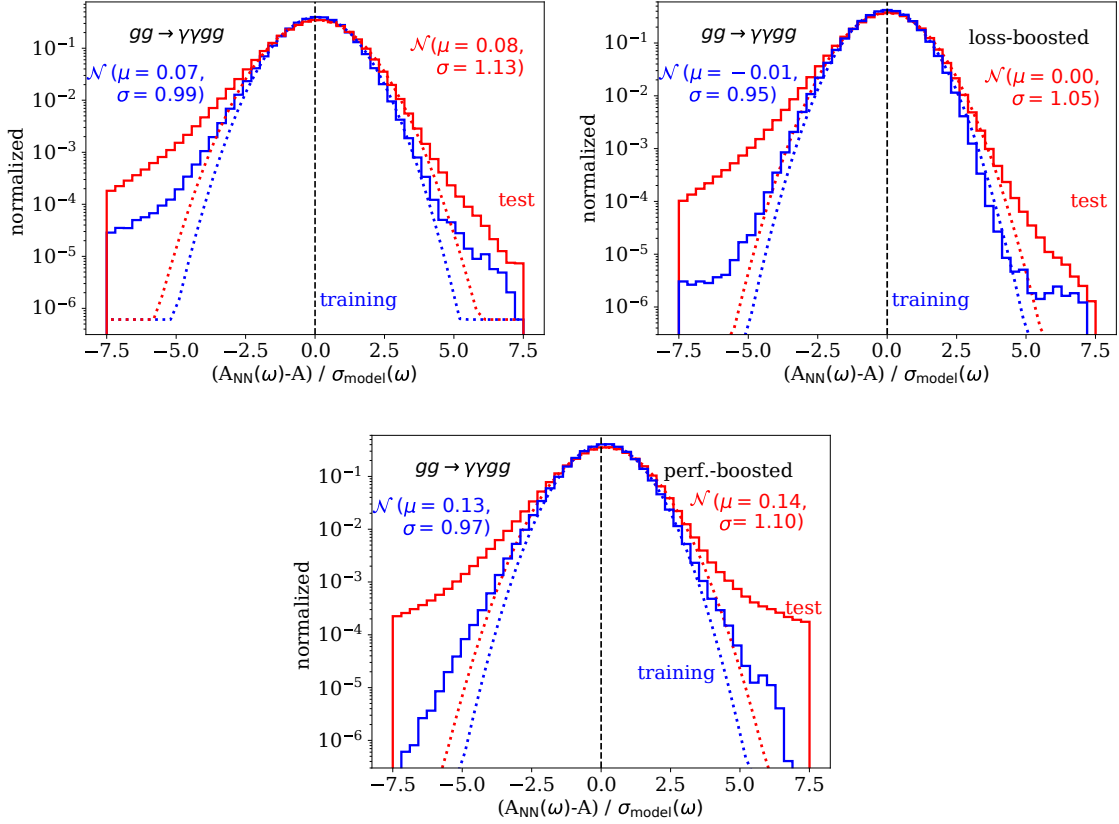

\centering
  \includegraphics[width=0.495\textwidth, page=4]{run224_pulls}
  \includegraphics[width=0.495\textwidth, page=4]{loss_boosted224_pulls}
  \includegraphics[width=0.495\textwidth, page=4]{process_boosted224_pulls}
  \caption{Pulls of the loss-bossted (left), and performance-boosted
    (right) BNN for the $(2 \to 4)$-process, defined in
    Eq.\eqref{eq:pull2} and evaluated on the training and test data.}
  \label{fig:2to4pulls}
\end{figure}

Next, we check the consistency of the network output by looking at the
$\omega$-dependent pull distribution defined Eq.\eqref{eq:pull2}. We
see that especially for large amplitudes the loss-boosting guarantees
a well-behaved, consistent network, while the additional process
boosting reverses some of the beneficial effects of the loss-boosting. 
This effect was already observed for the $(2\to3$)-process.

Finally, we show the 1-dimensional kinematic distributions for the
basic BNN and for the performance-boosted BNN. As for the $(2 \to
3)$-process the boosting step has a spectacula effect on the training
data in the poorly learned kinematic tails. After integrating over the
additional phase space directions this improvement translates well
into the test dataset, but at the expense of the uncertainty estimate
on the training data.

\begin{figure}[b!]
  \includegraphics[width=0.495\textwidth, page=4]{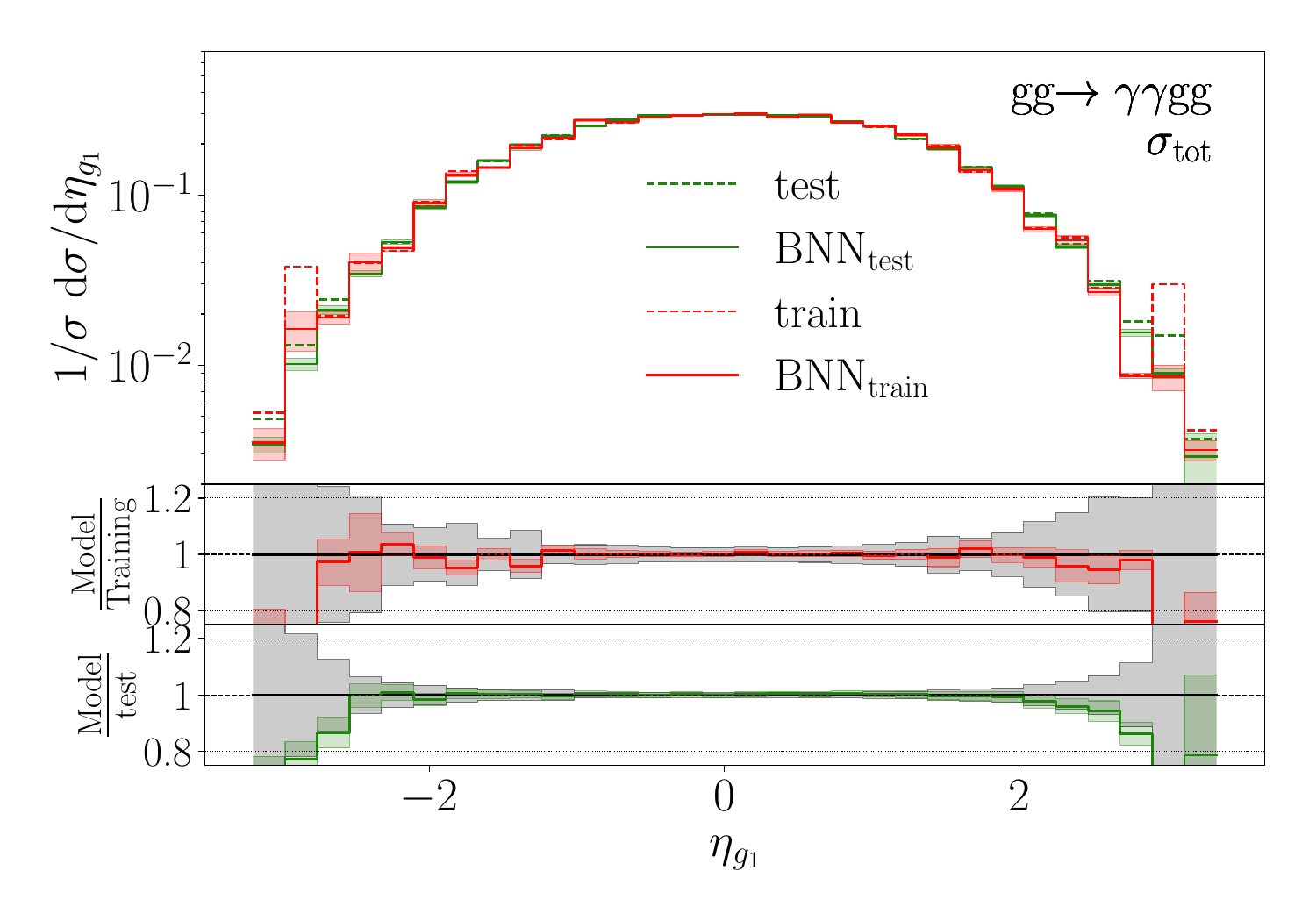}
  \includegraphics[width=0.495\textwidth, page=15]{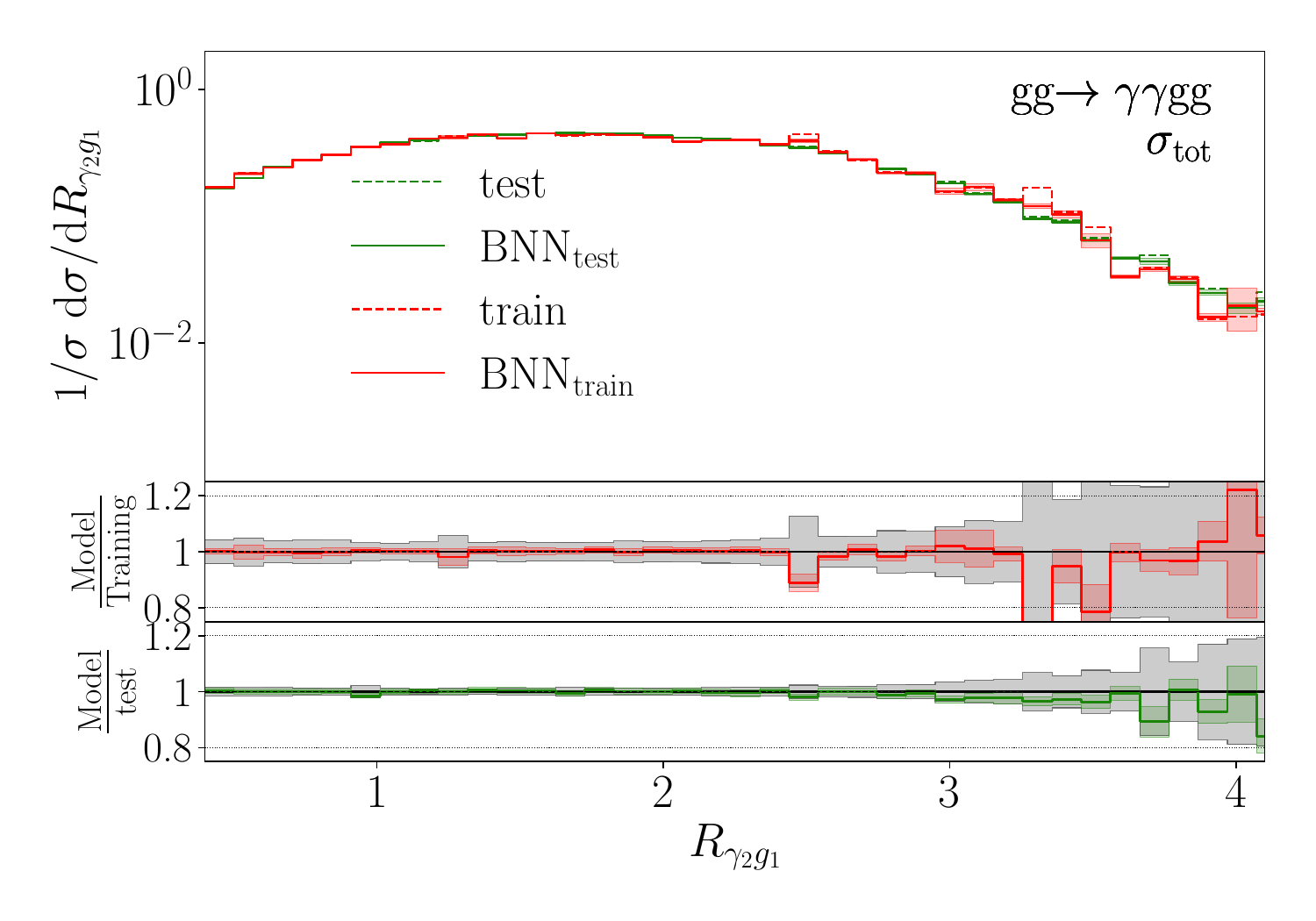}
  \includegraphics[width=0.495\textwidth, page=4]{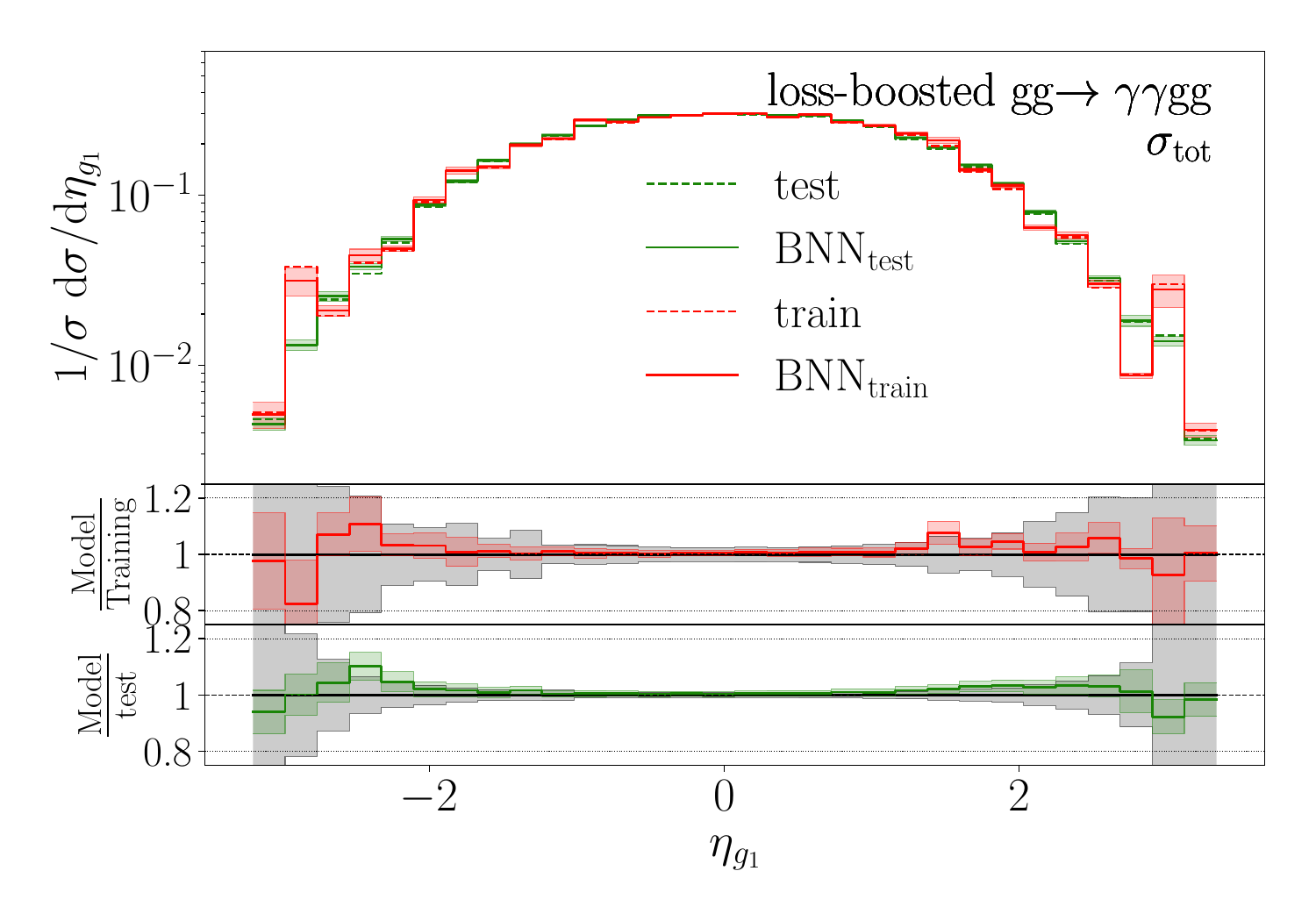}
  \includegraphics[width=0.495\textwidth, page=15]{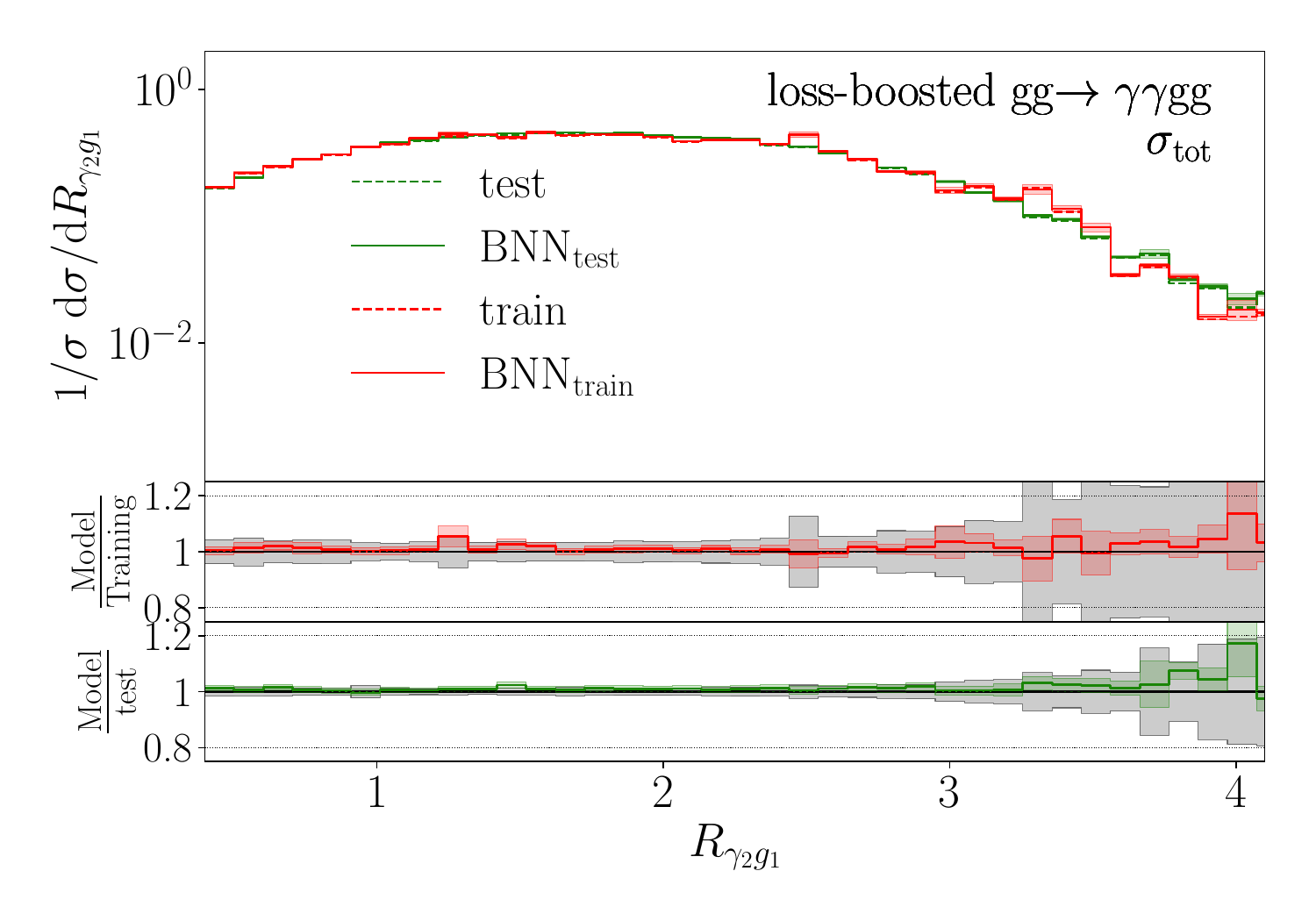}
  \includegraphics[width=0.495\textwidth, page=4]{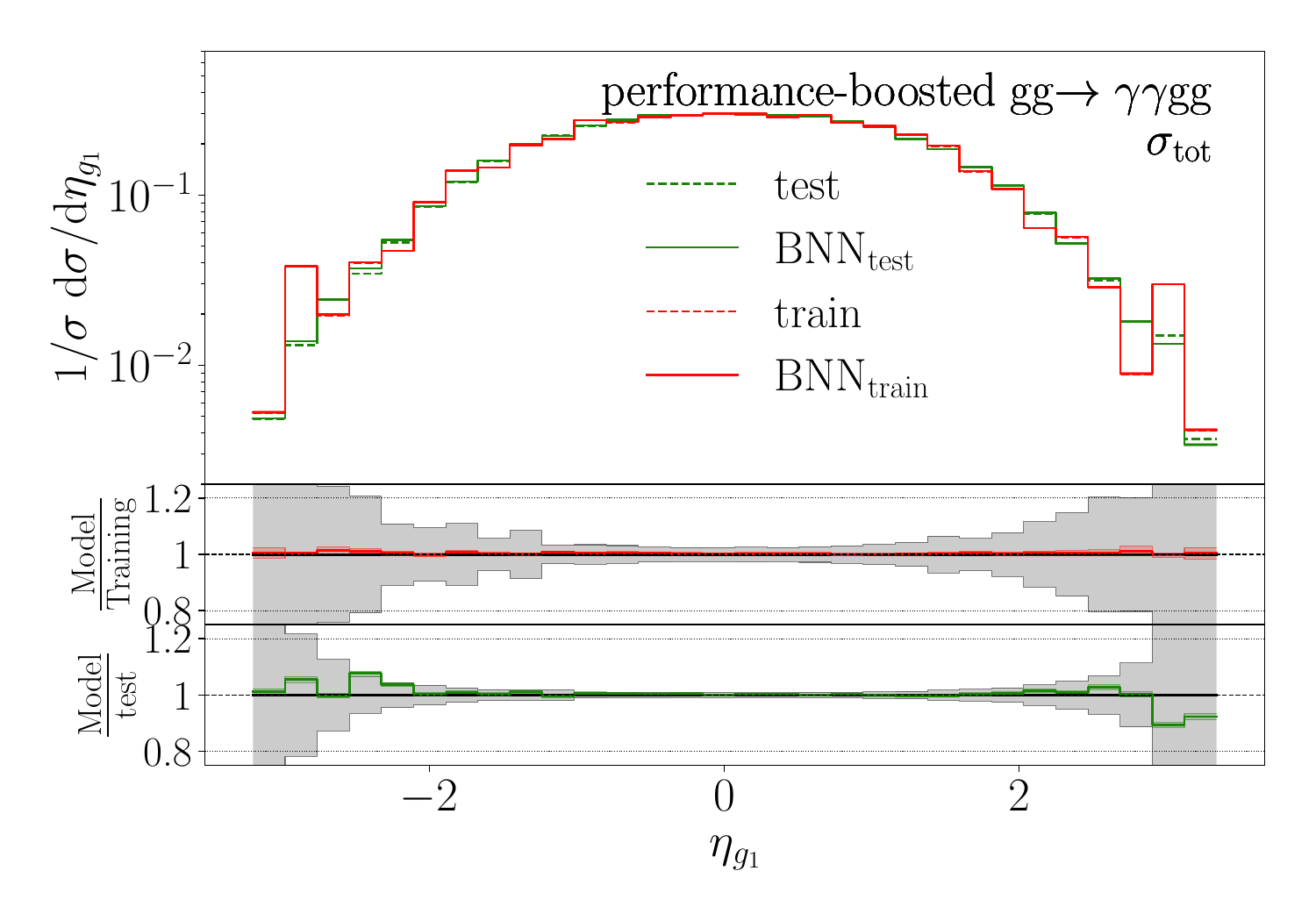}
  \includegraphics[width=0.495\textwidth, page=15]{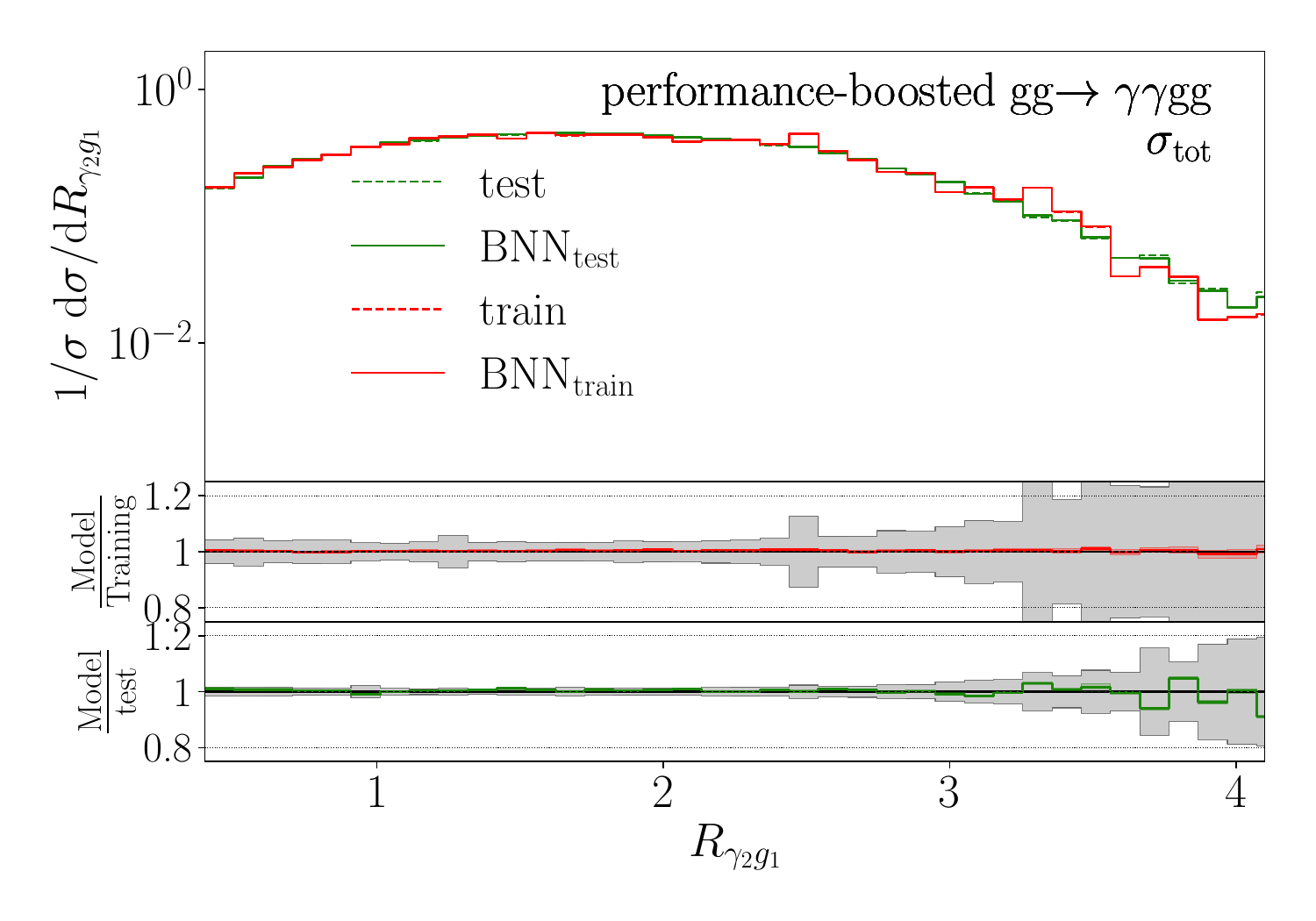}
  \caption{Kinematic distribution for the $(2 \to 4)$-process without
    boosting (upper), after loss boosting (center), and after process
    boosting (lower). The grey error bars in the lower panels indicate
    the statistical limitation of the training and test data.}
  \label{fig:2to4kin}
\end{figure}

\clearpage
\bibliographystyle{SciPost-bibstyle-arxiv}
\bibliography{paper}

\end{document}